\documentclass[11pt,letterpaper]{article}

\usepackage[T1]{fontenc}
\usepackage{lmodern}

\usepackage[utf8]{inputenc} 
\usepackage[T1]{fontenc}    
\usepackage[hidelinks]{hyperref}       
\usepackage{url}            
\usepackage{booktabs}       
\usepackage{amsfonts}       
\usepackage{nicefrac}       
\usepackage{microtype}      
\usepackage{natbib}
\usepackage{amsthm}
\usepackage{amsmath}
\usepackage{algcompatible}
\usepackage{algorithm}

\usepackage[noend]{algpseudocode}

\usepackage{booktabs}
\usepackage{enumerate}
\usepackage{graphicx}
\usepackage{amssymb}
\usepackage{latexsym}
\usepackage{epstopdf}
\usepackage{color}
\usepackage{xcolor}
\usepackage{bbm}
\usepackage{needspace}
\usepackage{color, colortbl}
\usepackage[english]{babel}

\usepackage{caption}
\usepackage{subcaption}
\usepackage{filecontents}
\usepackage{mathtools}
\usepackage{multirow}
\usepackage{tabularx}
\usepackage[margin=1.0in]{geometry}

\newtheorem{theorem}{Theorem}

\newtheorem{corollary}{Corollary}
\newtheorem{lemma}{Lemma}

\newtheorem{remark}{Remark}

\newtheorem{problem}{Problem}

\allowdisplaybreaks

\newcommand{\DDelta}{\mathbf{\Delta}}
\newcommand{\xx}{\mathbf{x}}
\newcommand{\uu}{\mathbf{u}}
\newcommand{\ww}{\mathbf{w}}
\newcommand{\bfeta}{\boldsymbol{\eta}}

\newcommand{\sA}{\mathbf{A}}
\newcommand{\sB}{\mathbf{B}}

\newcommand{\Phix}{\mathbf{\Phi}_x}
\newcommand{\Phiu}{\mathbf{\Phi}_u}
\newcommand{\KK}{\mathbf{K}}

\newcommand{\tildePhix}{\widetilde{\mathbf{\Phi}}_x}
\newcommand{\tildePhiu}{\widetilde{\mathbf{\Phi}}_u}
\newcommand{\tildeww}{\widetilde{\mathbf{w}}}

\newcommand{\conv}{\textrm{Conv}}
\newcommand{\bfSigma}{\mathbf{\Sigma}}
\newcommand{\facet}{\textrm{facet}}

\newcommand{\diag}{\textrm{diag}}
\newcommand{\blkdiag}{\textrm{blkdiag}}

\newcommand{\vertex}{\textrm{Vert}}

\title{Robust Model Predictive Control with Polytopic Model Uncertainty through System Level Synthesis } 

\author{Shaoru Chen, Victor M. Preciado, Manfred Morari, Nikolai Matni 
	\thanks{Shaoru Chen, Victor M. Preciado, Manfred Morari, Nikolai Matni are with the Department of Electrical and Systems Engineering, University of Pennsylvania, Philadelphia, PA, 19104, USA (e-mail: \{ srchen, preciado, morari, nmatni \}@seas.upenn.edu). Our codes are publicly available at \url{https://github.com/ShaoruChen/Polytopic-SLSMPC}.
	}
}

\date{}

\begin{document}
	
\pagestyle{plain}
\maketitle

\begin{abstract}                          
We propose a robust model predictive control (MPC) method for discrete-time linear systems with polytopic model uncertainty and additive disturbances. Optimizing over linear time-varying (LTV) state feedback controllers has been successfully used for robust MPC when only additive disturbances are present. However, it is challenging to design LTV state feedback controllers in the face of model uncertainty whose effects are difficult to bound. To address this issue, we propose a novel approach to over-approximate the effects of both model uncertainty and additive disturbances by a filtered additive disturbance signal. Using the System Level Synthesis framework, we jointly search for robust LTV state feedback controllers and the bounds on the effects of uncertainty online, which allows us to reduce the conservatism and minimize an upper bound on the worst-case cost in robust MPC. We provide a comprehensive numerical comparison of our method and representative robust MPC methods from the literature. Numerical examples demonstrate that our proposed method can significantly reduce the conservatism over a wide range of uncertainty parameters with comparable computational effort as the baseline methods.


\end{abstract}

\section{Introduction}
\label{sec:introduction}
In model predictive control (MPC), a finite time constrained optimal control problem (OCP) is solved at each time step and the first optimal control input is applied. When the system dynamics is uncertain, robust model predictive control explicitly takes the uncertainty into account by solving a robust OCP at each time step to guarantee that the state and control input constraints are robustly satisfied for the closed-loop system. Although dynamic programming~\citep[Chapter 15]{borrelli2017predictive} can exactly solve the robust OCP, it suffers from prohibitive computational complexity and is impractical to use. This has motivated the development of alternative solutions to the robust OCP which aim to reach a reasonable compromise between conservatism as measured by the size of the set of feasible initial states, and computational complexity. 

\textbf{Robust MPC with additive disturbances}: For uncertain linear time-invariant (LTI) systems with the only uncertainty arising from additive disturbances, a variety of closed-loop methods have been proposed to solve the robust OCP with feedback policies as decision variables. The use of feedback policies does not suffer from the conservatism of open-loop strategies in the presence of uncertainty~\citep{mayne2014model}. To maintain computational tractability, closed-loop methods only search over control policies that admit a finite-dimensional parameterization. Tube-based MPC is a widely used class of robust MPC methods where a tube (normally a sequence of polytopes) is constructed such that it contains all possible system trajectories under uncertainty. 
Different designs of tube MPC such as rigid~\citep{mayne2005robust}, homothetic~\citep{rakovic2012homothetic}, and elastic~\citep{rakovic2016elastic} tube MPC have been proposed with increasing flexibility in the parameterization of the tube and the associated control policy. One limitation of most tube-based MPC methods (see the introduction of~\cite{rakovic2012parameterized}) is the offline design of the tube cross-sections and the associated control law, e.g., by pre-fixing the shape of the cross-section and the feedback gain of the controller~\citep{rakovic2012homothetic}, which can be conservative. 

Compared with tube-based MPC, state feedback MPC~\citep{lofberg2003approximations, goulart2006optimization} optimizes the state feedback or the equivalent disturbance feedback gains online~\citep{goulart2006optimization}, resulting in reduced conservatism at the cost of increased computational complexity. In fact, the number of decision variables scales linearly in the horizon for tube-based MPC using an offline design, and quadratically in the horizon for state/disturbance feedback MPC. A conservatism improvement together with the quadratic complexity is also achieved by the parameterized tube MPC~\citep{rakovic2012parameterized} which adopts online optimization over both the tube and the control policies. \cite{sieber2021system} propose to search over state feedback gains online through the framework of System Level Synthesis (SLS)~\citep{anderson2019system} with structural restrictions such that the number of variables is linear in the horizon.

\textbf{Robust MPC with model uncertainty}: When model uncertainty is present, robust MPC becomes significantly more challenging since the perturbation to the nominal predicted trajectory directly depends on the states and control inputs and hence on both the uncertainty and the controller parameters.

For polytopic model uncertainty, robust MPC methods based on linear matrix inequalities (LMI) are presented in~\cite{kothare1996robust, schuurmans2000robust, kouvaritakis2000efficient, munoz2015robust} where affine state feedback policies for a min-max cost function over an infinite horizon are considered. The LMI-based methods do not require much offline parameter design, but their computational complexity tends to scale badly with the system dimension because of the use of semidefinite programming~\citep{gesser2018robust}. In addition, the ellipsoidal inner approximation of polyhedral state and control input constraints~\citep{kothare1996robust} can be conservative. 

The application of tube-based MPC to handle model uncertainty can be found in~\citep{langson2004robust, fleming2014robust, lu2019robust, kohler2019linear}. \citet{langson2004robust} utilize a homothetic tube to bound the system trajectories and parameterize a feedback control policy associated with each vertex of the tube cross-section. \citet{fleming2014robust, lu2019robust, kohler2019linear} apply an offline-designed tube and a control policy $u_t = K x_t + v_t$ with a pre-stabilizing feedback gain $K$ to guarantee robust constraint satisfaction. Similar to tube MPC in the disturbance-only scenario, the offline design of either the tube or the associated control policy can be conservative. 

Recent works by~\cite{bujarbaruah2021simple, bujarbaruah2022robust} use LTV state feedback controllers for robust MPC under model uncertainty. Because the LTV state feedback controller has more degrees of freedom, the analysis of the effects of uncertainty is more complex and guaranteeing robust constraint satisfaction is more challenging. Therefore, a simplified, conservative uncertainty over-approximation is needed to solve the robust OCP in a numerically efficient manner. \citet{bujarbaruah2021simple} adopt a simple approach that globally over-approximates the effects of model uncertainty by additive disturbances with fixed norm bounds and reduces the robust OCP to the disturbance-only case discussed above.  Although easily implementable, this method suffers from the inherent conservatism induced by the global uncertainty over-approximation which fails to utilize the structure of the model uncertainty. In \cite{bujarbaruah2022robust}, the effects of model uncertainty are first bounded offline as a function of the control inputs, and an LTV state feedback controller is optimized online afterward. We denote this class of methods as \emph{uncertainty over-approximation-based} MPC to highlight the main challenge of bounding the uncertainty when an LTV state feedback controller is considered. 

\subsection{Contributions}
In this work, we propose a robust MPC method which optimizes over LTV state feedback controllers online for systems subject to polytopic model uncertainty and additive disturbances. Inspired by~\cite{bujarbaruah2021simple}, we describe the uncertain system dynamics as the sum of the nominal dynamics and the lumped uncertainty that captures the deviation from the nominal predicted state at each time instant. Unlike~\cite{bujarbaruah2021simple} who simply over-approximate the lumped uncertainty uniformly by a sufficiently large additive disturbance signal, our method exactly characterizes how the lumped uncertainty depends on the model uncertainty and controller parameters by leveraging the framework of System Level Synthesis, which establishes the equivalence between the optimization of LTV state feedback controllers and closed-loop system responses. To guarantee robust constraint satisfaction, we propose a novel method that employs a parameterized filtered additive disturbance signal to over-approximate the lumped uncertainty at each time step. Importantly, our method retains the dependence of the uncertainty over-approximation on the controller parameters and thus allows joint optimization of the uncertainty over-approximation parameters and the controller in a convex manner.

We denote the proposed  method as \emph{SLS MPC} since our robust OCP formulation is derived in the space of system responses made possible by SLS. By construction, SLS MPC considers a more flexible controller class than the tube-based MPC methods~\citep{fleming2014robust, lu2019robust, kohler2019linear} which use a fixed pre-stabilizing controller. Compared with methods~\cite{bujarbaruah2021simple, bujarbaruah2022robust} that also apply LTV state feedback controllers, SLS MPC handles the uncertainty over-approximation problem using a novel approach, the effectiveness of which is demonstrated by numerical examples. 
Our contributions are summarized as follows.

\begin{enumerate}
	\item We propose a novel robust MPC method, SLS MPC, for uncertain LTI systems with polytopic model uncertainty and additive disturbances. The method solves the robust OCP by jointly optimizing LTV state feedback controllers and uncertainty over-approximation parameters through a convex quadratic program. In addition to the typical nominal cost~\citep{langson2004robust, fleming2014robust,kohler2019linear, bujarbaruah2022robust},	our method allows minimizing upper bounds over the worst-case costs with respect to the model uncertainty in robust MPC.
	\item We provide a comprehensive comparison of tube-based~\citep{langson2004robust, lorenzen2019robust, kohler2019linear, lu2019robust} and uncertainty over-approximation-based~\citep{bujarbaruah2021simple, bujarbaruah2022robust} robust MPC methods, including SLS MPC, through numerical examples. We demonstrate that SLS MPC can significantly reduce conservatism compared with all other methods over a wide range of uncertainty parameters. 
\end{enumerate}


The rest of the paper is organized as follows. The uncertain LTI system and the problem formulation are introduced in Section~\ref{sec:formulation}. In Section~\ref{sec:lumped_uncertainty}, we introduce System Level Synthesis and use it to characterize the effects of model uncertainty, which are over-approximated in Section~\ref{sec:over_approx} by a filtered disturbance signal. Section~\ref{sec:convex_OCP} presents the proposed SLS MPC method, derives upper bounds on the worst-case costs, and discusses the closed-loop properties of SLS MPC. Extensive simulation is given in Section~\ref{sec:simulation} to compare SLS MPC with representative baseline methods. Section~\ref{sec:conclusion} concludes the paper. 

\textbf{Notation}:  For a dynamical system, we denote the system state at time $k$ by $x(k)$ and the $t$-step prediction of the state in an MPC loop by $x_t$. For two vectors $x$ and $y$, $x \leq y$ denotes element-wise comparison. For a symmetric matrix $Q$, $Q \succeq 0$ denotes that $Q$ is positive semidefinite. The notation $x_{i:j}$ is shorthand for the set $\{x_i, x_{i+1}, \cdots, x_j\}$.  For a vector $d \in \mathbb{R}^n$, $S = \diag(d)$ denotes a $n \times n$ dimensional diagonal matrix with $d$ being the diagonal vector. For a sequence of matrices $S_1, \cdots, S_N$, $S = \text{blkdiag}(S_1, \cdots, S_N)$ denotes that $S$ is a block diagonal matrix whose diagonal blocks are $S_1, \cdots, S_N$ arranged in the order. We represent a linear, causal operator $\mathbf{R}$ defined over a horizon $T$ by the block-lower-triangular matrix
\begin{equation} \label{eq:BLT}
	\mathbf{R} = \begin{bmatrix}
		R^{0,0} & \ & \ & \ \\
		R^{1,1} & R^{1,0} & \ & \ \\
		\vdots & \ddots & \ddots & \ \\
		R^{T,T} & \cdots &R^{T,1} & R^{T,0}
	\end{bmatrix}
\end{equation} 
where $R^{i,j} \in \mathbb{R}^{p \times q}$ is a matrix of compatible dimension. The set of such matrices is denoted by $\mathcal{L}_{TV}^{T, p \times q}$ and we will drop the superscript $T$ or $p \times q$ when it is clear from the context. Let $\mathbf{R}(i,:)$ denote the $i$-th block row of $\mathbf{R}$, and $\mathbf{R}(:,j)$ denote the $j$-th block column of $\mathbf{R}$, both indexing from $0$\footnotemark. The support function of a non-empty, closed convex set $\mathcal{X}$ is defined as $h_\mathcal{X}(c) = \max \{c^\top x \text{ s.t. } x \in \mathcal{X}\}$. \footnotetext{In this paper, we refer to a block matrix in a block-lower-triangular matrix $\mathbf{R}$ using its superscripts shown in Eq.~\eqref{eq:BLT}. If we let $\mathbf{R}(i,j)$ denote the block matrix in the $i$-th row and $j$-th column, then we have $\mathbf{R}(i, j) = R^{i, i-j}$ with $(i, j)$ indexing from $0$. }

\section{Problem formulation}
\label{sec:formulation}
Consider a discrete-time linear system with uncertain dynamics
\begin{equation} \label{eq:dyn}
x(k+1) = (\hat{A} + \Delta_A) x(k) + (\hat{B} + \Delta_B) u(k) + w(k)
\end{equation}
for $k \geq 0$ where $x(k) \in \mathbb{R}^{n_x}$ is the state, $u(k)\in\mathbb{R}^{n_u}$ is the control input, $w(k) \in\mathbb{R}^{n_x}$ is the additive disturbance, and $(\hat{A}, \hat{B})$ are the known nominal dynamics. The matrices $(\Delta_A, \Delta_B)$ denote additive model uncertainty and belong to a polytopic set comprising $M$ vertices
\begin{equation} \label{eq:polytopic_uncertainty}
(\Delta_A, \Delta_B) \in \conv\{(\Delta_{A,1}, \Delta_{B,1}), \cdots,(\Delta_{A,M}, \Delta_{B,M}) \}
\end{equation}
with $\conv$ denoting the convex hull. For simplicity, we denote the polytopic model uncertainty set defined in~\eqref{eq:polytopic_uncertainty} as $\mathcal{P}$~\footnote{While in this work we focus on solving the robust MPC problem, our proposed method can be easily incorporated into the adaptive MPC framework where the uncertainty parameters $(\Delta_A, \Delta_B)$ are assumed time-invariant and the polytopic model uncertainty set $\mathcal{P}$ is updated online as shown by~\citet{lu2019robust} and \citet{kohler2019linear}. }. The additive disturbance $w(k)$ is assumed to lie in a polytopic set, i.e, $w(k) \in \mathcal{W}$ where $\mathcal{W}$ is a polytope.

In robust MPC, at each time instant $k$, we solve a finite time robust OCP which aims to synthesize a feedback controller that guarantees the robust satisfaction of state and input constraints. The first control input in the optimal solution is applied to drive the system to the next state, and this process is repeated. In this paper, we focus on solving the following robust OCP with polytopic model uncertainty and additive disturbances. 

\begin{problem}\label{prob:robustmpc}
	Solve the following finite time constrained robust OCP with horizon $T$:
	\begin{equation} \label{eq:robustOCP}
	\begin{aligned}
	 J_T^* &= min_{\mathbf{\pi} }  \quad J_T(\pi)\\
	\text{s.t.} &  \quad x_{t + 1} = (\hat{A} + \Delta_A) x_{t} + (\hat{B} + \Delta_B) u_{t} + w_t \\
	& \quad u_t = \pi_t(x_{0:t}) \\
	& \quad x_t \in \mathcal{X}, u_t \in \mathcal{U}, x_T \in \mathcal{X}_T, t = 0, 1, \cdots, T - 1 \\
	& \quad \forall (\Delta_A, \Delta_B) \in \mathcal{P}, \forall w_t \in \mathcal{W}, t = 0, 1, \cdots, T - 1 \\
	& \quad  x_0 = x(k)
	\end{aligned}
	\end{equation}
	where the search is over causal LTV state feedback control policies $\mathbf{\pi} = \pi_{0:T-1}$ and $J_T(\pi)$ denotes the nominal cost function as is typical in the robust MPC literature~\footnote{The worst-case cost function is considered in Section~\ref{sec:convex_OCP_formulation}. }:
	\begin{equation} \label{eq:nominalcost}
		\begin{array}{rl}
			J_{T}(\pi) & =  \sum_{t = 0}^{T - 1} (\hat{x}_t^\top Q \hat{x}_t + u_t^\top R u_t ) + \hat{x}_T^\top Q_T \hat{x}_T \\
			\text{s.t.} \ & \hat{x}_{t+1} = \hat{A} \hat{x}_t + \hat{B} u_t,  u_t = \pi_t(\hat{x}_{0:t}) \\
			& \hat{x}_0 = x(k), \quad \forall t = 0, 1, \cdots, T-1,
		\end{array}
	\end{equation}
	where $\hat{x}_{0:T}$ denotes the nominal trajectory, and $Q \succeq 0, R \succ 0$, $Q_T \succeq 0$ denote the state, input, and terminal weight matrices, respectively.
	The sets $\mathcal{X}, \mathcal{U}$, and $\mathcal{X}_T$ are the polytopic state, input, and terminal constraints, defined as
	\begin{equation*} \label{eq:constraints}
	\begin{aligned}
	&\mathcal{X} = \lbrace x \in \mathbb{R}^{n_x} \mid F_x x \leq b_x \rbrace, \  
	\mathcal{U} = \lbrace u \in \mathbb{R}^{n_u} \mid F_u u \leq b_u \rbrace,\\
	&\mathcal{X}_T = \lbrace x \in \mathbb{R}^{n_x} \mid F_T x \leq b_T \rbrace.
	\end{aligned}
	\end{equation*}
	We assume that the sets $\mathcal{X}, \mathcal{U}$, and $\mathcal{X}_T$ are compact and contain the origin in their interior. 
\end{problem}

As shown by~\citet[Chapter 15]{borrelli2017predictive}, the robust OCP~\eqref{eq:robustOCP} can be solved exactly by dynamic programming which has a high computational cost. For numerical tractability, a finite-dimensional, parameterized controller $\pi_t$ has to be considered in~\eqref{eq:robustOCP} and the state and input constraints are tightened to guarantee robust constraint satisfaction. 
In this work, we consider searching over causal LTV state feedback controllers represented by $u_t = \sum_{i= 0}^t K^{t, t-i} x_i$ for $t = 0, \cdots, T-1$ with $K^{t,t-i}$ being our design parameters. Our approach to Problem~\eqref{eq:robustOCP} relies on over-approximating the effects of uncertainties in the space of closed-loop system responses as shown in the following sections. 


\section{Characterization of effects of uncertainty}
\label{sec:lumped_uncertainty}
In the robust OCP, the system dynamics can be decomposed as the sum of nominal dynamics and uncertainty-related terms:
\begin{equation} \label{eq:system_dyn}
	\begin{aligned}
	x_{t+1} &= \hat{A} x_t + \hat{B} u_t + \underbrace{\Delta_A x_t + \Delta_B u_t + w_t}_{\eta_t} \\
	& =  \hat{A} x_t + \hat{B} u_t + \eta_t
	\end{aligned}
\end{equation}
where we define $\eta_t := \Delta_A x_t + \Delta_B u_t + w_t$ as the \emph{lumped uncertainty} which models the perturbation to the nominal dynamics at each time step. To find a feasible solution to~\eqref{eq:robustOCP} with minimal conservatism, a proper characterization of $\eta_t$ is required. However, this is challenging since the lumped uncertainty depends on both the uncertainty parameters and the feedback controller to be designed. In this section, we show that SLS allows us to \emph{exactly} characterize the dynamics of the lumped uncertainty $\eta_t$ as a function of both the uncertainty and the controller parameters. 

\subsection{Finite-horizon System Level Synthesis}
To apply SLS, we first stack all relevant state, control input, and uncertainty variables over horizon $T$ as 
\begin{equation}\label{eq:signal_def}
\begin{aligned}
\xx = [x_0^\top \ \cdots \ x_T^\top]^\top, \quad \uu = [u_0^\top \ \cdots \ u_T^\top]^\top, \quad \bfeta = [x_0^\top \ \eta_0^\top \ \cdots \ \eta_{T-1}^\top]^\top, \quad  \ww = [x_0^\top \ w_0^\top \ \cdots \ w_{T-1}^\top]^\top.
\end{aligned}
\end{equation}
Note that the initial state $x_0$ is set as the first component in $\bfeta$ and $\ww$, and $x_0$ can be interpreted as a known disturbance from the origin in this case. The vectors in~\eqref{eq:signal_def} can be interpreted as finite horizon signals. The parameterization of the LTV state feedback controller $\KK \in \mathcal{L}_{TV}^{T, n_u \times n_x}$ is represented by the block-lower-triangular matrix~\eqref{eq:BLT} with entries $K^{t, t-i}$. The controller $\KK$ can be interpreted as a time-varying linear operator and the state feedback controller is given by $\uu = \KK \xx$. Similarly, we stack the dynamics matrices and uncertainty matrices as 
\begin{equation} \label{eq:matrices_stack}
\begin{aligned}
	&\mathbf{\hat{A}} = \blkdiag(\hat{A}, \cdots, \hat{A}), \quad \mathbf{\hat{B}} = \blkdiag(\hat{B}, \cdots, \hat{B}), \\
	&\mathbf{\Delta}_A \!=\! \blkdiag(\Delta_A, \! \cdots \!, \Delta_A),\! \mathbf{\Delta}_B \!=\! \blkdiag(\Delta_B, \! \cdots \!, \Delta_B),
\end{aligned}
\end{equation}
which all belong to $\mathcal{L}_{TV}^T$. With this compact notation, the open-loop dynamics of the system~\eqref{eq:system_dyn} can be written as
\begin{align}
	&\xx = Z \hat{\sA}\xx + Z\hat{\sB} \uu + \bfeta,
\end{align}
and the closed-loop dynamics under $\uu = \KK \xx$ follows as 
\begin{equation} \label{eq:cl_dyn}
 \xx = Z (\hat{\sA} + \hat{\sB} \KK) \xx + \bfeta,
\end{equation}
where $Z$ is a block-downshift operator with the first block sub-diagonal filled with identity matrices and zeros everywhere else. Note that the lumped uncertainty $\bfeta$ depends on $\KK$ and $\xx$ as will be shown in the next subsection. 

From~\eqref{eq:cl_dyn}, the mapping from the lumped uncertainty to the state and control input under the feedback controller is given by 
\begin{equation} \label{eq:explicit_map}
\begin{bmatrix}
\xx \\ \uu 
\end{bmatrix} = \begin{bmatrix}
(I - Z (\hat{\sA} + \hat{\sB} \KK))^{-1} \\
\KK (I - Z (\hat{\sA} + \hat{\sB} \KK))^{-1}
\end{bmatrix} \bfeta.
\end{equation}
Because of the block-downshift operator $Z$, we know that the matrix inverse in~\eqref{eq:explicit_map} exists and has a block-lower-triangular structure. The maps from $\bfeta$ to $(\xx, \uu)$ in~\eqref{eq:explicit_map} are called \emph{system responses}, and we denote them by $\Phix \in \mathcal{L}_{TV}^{T, n_x \times n_x}$, $\Phiu \in \mathcal{L}_{TV}^{T, n_u \times n_x}$ such that
\begin{equation} \label{eq:system_response}
\begin{bmatrix}
\xx \\ \uu 
\end{bmatrix} = \begin{bmatrix}
\Phix \\
\Phiu
\end{bmatrix} \bfeta.
\end{equation}
The relationship between the LTV state feedback controller $\KK$ and the system responses $(\Phix, \Phiu)$ can be written explicitly as:
\begin{equation} \label{eq:K_to_Phi}
	\begin{aligned}
		&\Phix = (I - Z (\hat{\sA} + \hat{\sB} \KK))^{-1}, \\
		&\Phiu = \KK (I - Z (\hat{\sA} + \hat{\sB} \KK))^{-1}.
	\end{aligned}
\end{equation}
The following theorem allows us to transform the design of the feedback controller $\KK$ into the design of the system responses $\{ \Phix, \Phiu \}$ without explicitly using the nonlinear map~\eqref{eq:explicit_map}.
\begin{theorem}~\citep[Theorem 2.1]{anderson2019system} 
	\label{thm:SLS}
	Over the horizon $ t = 0, 1, \cdots, T$, for the system dynamics~\eqref{eq:system_dyn} with the block-lower-triangular state feedback control law $\KK \in \mathcal{L}_{TV}^{T, n_u \times n_x}$ defining the control action as $\uu = \KK \xx$, we have:
	\begin{enumerate}
		\item The affine subspace defined by 
		\begin{equation} \label{eq:affine_constr}
		\begin{bmatrix} I - Z \hat{\sA} & -Z \hat{\sB} \end{bmatrix} 
		\begin{bmatrix} \Phix \\ \Phiu \end{bmatrix} = I, \ \Phix, \Phiu \in \mathcal{L}_{TV}^T
		\end{equation}
		parameterizes all possible system responses~\eqref{eq:system_response}.
		\item For any block-lower-triangular matrices $\lbrace \Phix, \Phiu \rbrace \in \mathcal{L}_{TV}^T$ satisfying~\eqref{eq:affine_constr}, the controller $\KK = \Phiu \Phix^{-1} \in \mathcal{L}_{TV}^T$ achieves the desired responses~\eqref{eq:system_response}. 
	\end{enumerate}
\end{theorem}

Theorem~\ref{thm:SLS} shows the equivalence between system responses and LTV state feedback controllers through the affine constraint~\eqref{eq:affine_constr}. Therefore, we can refer to either $\KK$ or $\{ \Phix, \Phiu \}$ as the controller parameters. With constraint~\eqref{eq:affine_constr}, an optimization problem originally in $\KK$ can be transformed into one on $\{ \Phix, \Phiu \}$.  Such a transformation may result in a convex problem in $\{ \Phix, \Phiu \}$ while the original one is not. It provides a direct description of the effects of the lumped uncertainty $\bfeta$ on the states and control inputs. More importantly, as will be shown next, this transformation can reveal additional structural properties of the robust OCP in the space of system responses that can be exploited to reduce the conservatism of the solution.

\subsection{Dynamics of lumped uncertainty} 
\label{sec:eta_dynamics}
By the definition of lumped uncertainty~\eqref{eq:system_dyn} and the compact notations from~\eqref{eq:signal_def}, \eqref{eq:matrices_stack}, we have
\begin{equation} 
\bfeta = Z \begin{bmatrix}
\DDelta_A & \DDelta_B 
\end{bmatrix} \begin{bmatrix}
\xx \\ \uu
\end{bmatrix} + \ww.
\end{equation}
The system responses~\eqref{eq:system_response} allow an explicit characterization of the dynamics of $\eta_t$ under the controller $\KK$ as
\begin{equation} \label{eq:eta_dynamics}
\bfeta = Z \begin{bmatrix}
\DDelta_A & \DDelta_B 
\end{bmatrix} \begin{bmatrix}
\Phix \\ \Phiu
\end{bmatrix} \bfeta + \ww
\end{equation}
which can be decomposed into the following set of equations~\footnote{We adopt the convention that when $t=0$, the summation terms in~\eqref{eq:decomposed_dynamics} vanish. }
\begin{equation}\label{eq:decomposed_dynamics}
\begin{aligned}
\eta_t & = \Delta_A (\Phi_x^{t, t} x_0 + \sum_{i=1}^{t} \Phi_x^{t, t-i} \eta_{i-1} ) + \Delta_B (\Phi_u^{t, t} x_0 + \sum_{i=1}^{t} \Phi_u^{t, t-i} \eta_{i-1} ) + w_t
\end{aligned}
\end{equation}
for $t = 0, \cdots, T-1$. Since $\eta_t$ only depends on $\eta_i$ with $i \leq t-1$, i.e., the lumped uncertainty before time $t$,  it follows that the values of $\eta_t$ are \emph{uniquely} determined by the uncertainty parameters $\{ \DDelta_A, \DDelta_B, \ww \}$ and the closed-loop system responses $\{\Phix, \Phiu\}$. Therefore, we can treat $\bfeta$ as a function of the uncertainty and controller parameters. 


When $(\Delta_A, \Delta_B)$ and $w_t$ are unknown, the values of $\eta_t$ become uncertain. Under the uncertainty assumptions in Section~\ref{sec:formulation}, we denote by $\mathcal{R}(\bfeta; \{\Phix, \Phiu \})$ the set of all possible values of $\bfeta$ under a given controller $\KK = \Phiu \Phix^{-1}$:
\begin{equation}
	\begin{aligned}
	\mathcal{R}(\bfeta; \{\Phix, \Phiu \})  := & \{ \bfeta \mid  \exists  w_t \in \mathcal{W}, 0 \leq t \leq T-1,  \text{and } (\Delta_A, \Delta_B) \in \mathcal{P}, \text{ s.t. }  \eqref{eq:decomposed_dynamics} \text{ holds} \}.
	\end{aligned}
\end{equation}

We call $\mathcal{R}(\bfeta; \{\Phix, \Phiu \})$ the reachable set of $\bfeta$ under the controller $\KK = \Phiu \Phix^{-1}$. The second argument $\{ \Phix, \Phiu\}$ highlights the dependence of the reachable set on the controller parameters. 
Despite being exact, Eq.~\eqref{eq:eta_dynamics} or the derived reachable set $\mathcal{R}(\bfeta; \{\Phix, \Phiu \})$ is too complex to use for directly solving the robust OCP~\eqref{eq:robustOCP}. To address this issue, in the next section, we over-approximate $\mathcal{R}(\bfeta;\{\Phix, \Phiu\})$ by an uncertainty set with a simpler structure while retaining the dependence on the system responses $\{ \Phix, \Phiu \}$.

\section{Uncertainty over-approximation}
\label{sec:over_approx}
To over-approximate $\mathcal{R}(\bfeta;\{\Phix, \Phiu\})$, we design a disturbance signal represented by $\bfSigma \tildeww$ where $\bfSigma \in \mathcal{L}_{TV}^{T, n_x \times n_x}$ is a filter operating on a normalized unit norm-bounded virtual disturbance signal
\begin{equation}\label{eq:unit_norm}
\tildeww = [x_0^\top \ \tilde{w}_0^\top \ \cdots \ \tilde{w}_{T-1}^\top]^\top \textrm{ with } \lVert \tilde{w}_t \rVert_\infty \leq 1.
\end{equation}
We denote the set of $\tildeww$ satisfying the unit norm bound constraint~\eqref{eq:unit_norm} as $\mathcal{W}_{\tildeww}$, and define the reachable set of the filtered disturbance $\bfSigma \tildeww$ as
\begin{equation}
\mathcal{R}(\bfSigma \tildeww) := \{ \bfSigma \tildeww \mid \tildeww \in \mathcal{W}_{\tildeww}  \}
\end{equation} 
When parameterizing $\bfSigma$, we require $\Sigma^{0,0} = I$ such that the first component of $\bfSigma \tildeww$ is $x_0$. Our goal is to over-approximate the reachable set of the lumped uncertainty $\mathcal{R}(\bfeta;\{\Phix, \Phiu\})$ by that of the filtered disturbances $\mathcal{R}(\bfSigma \tildeww)$, i.e., $\mathcal{R}(\bfeta;\{\Phix, \Phiu\}) \subseteq \mathcal{R}(\bfSigma \tildeww)$, such that it suffices to consider the dynamical system 
\begin{equation}\label{eq:surrogate_dynamics}
\xx = Z\hat{\sA} \xx + Z \hat{\sB} \uu + \bfSigma \tildeww
\end{equation}
with the surrogate disturbances $\bfSigma \tildeww$ for solving the robust OCP. The unit norm-bounded assumption on the virtual disturbances $\tilde{w}_t$ simplifies the constraint tightening of the robust OCP~\eqref{eq:robustOCP}, while the filter $\bfSigma$, which is our design parameter, controls the complexity of $\mathcal{R}(\bfSigma \tildeww)$ such that it can over-approximate $\mathcal{R}(\bfeta;\{\Phix, \Phiu\})$ with minimal conservatism. Next, we discuss the parameterization of the filter $\bfSigma$ and formulate a set of linear constraints on $\bfSigma$ and $\{\Phix, \Phiu\}$ such that $\mathcal{R}(\bfeta;\{\Phix, \Phiu\}) \subseteq \mathcal{R}(\bfSigma \tildeww)$ holds.

\subsection{Parameterization of the filter}
\label{sec:filter_parameterization}
\begin{figure}
\centering
\includegraphics[width = 0.6 \textwidth]{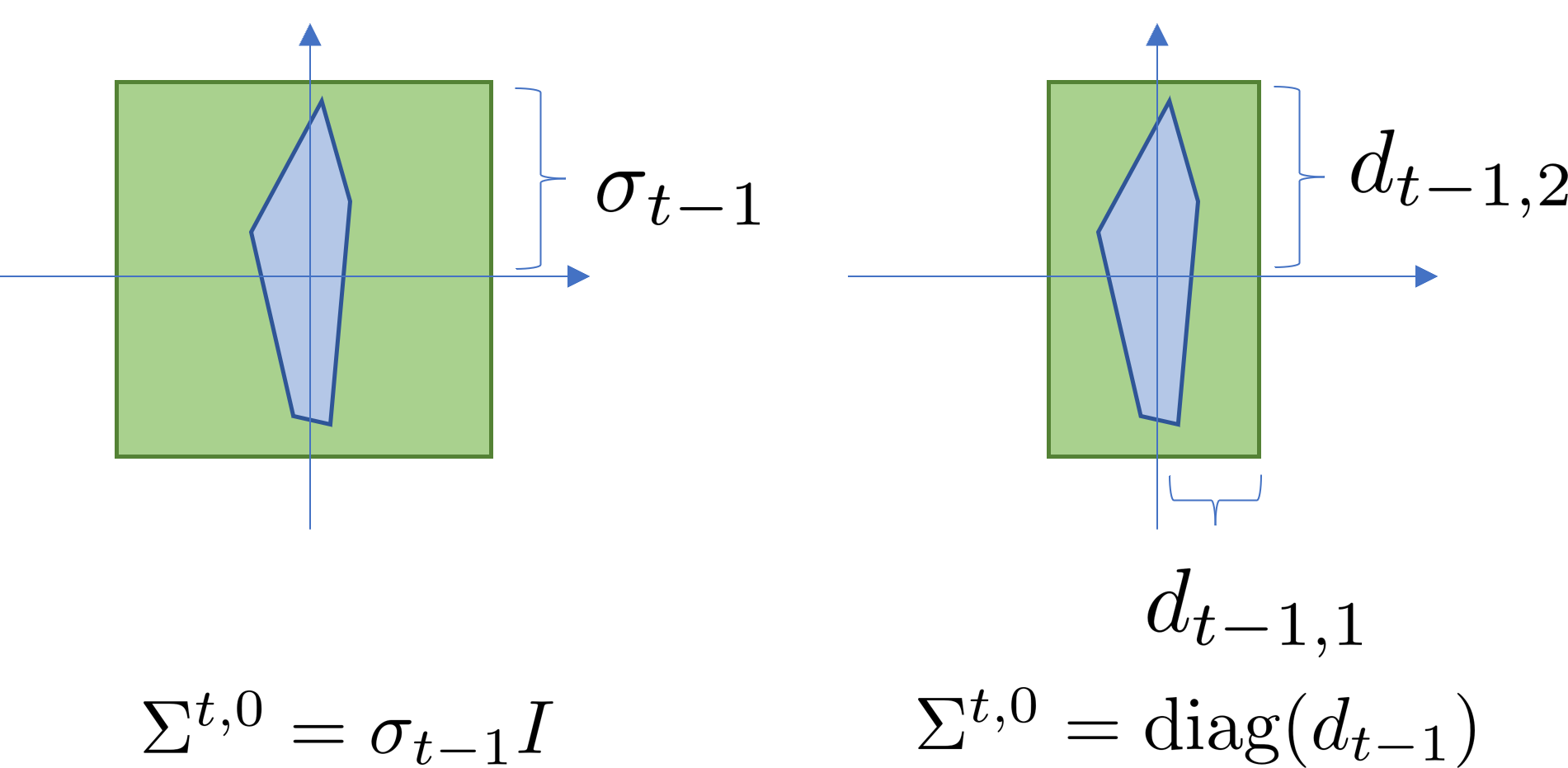}
\caption{Blue: reachable set of the lumped uncertainty $\eta_t$. Green: uncertainty over-approximation by the filtered disturbance signal $\sigma_t \tilde{w}_t$ with the block diagonal parameterization of the filter $\bfSigma$. The norm ball (left) and hyperrectangle (right) over-approximations are demonstrated. }
\label{fig:ball_bound}
\end{figure}

To motivate the use of the filtered disturbance signal $\bfSigma \tildeww$ for lumped uncertainty over-approximation, we first consider two special block diagonal parameterizations of $\bfSigma$, i.e., only the matrices $\Sigma^{t,0}$ on the diagonal of $\bfSigma \in \mathcal{L}_{TV}$ are non-zero. Note that we always have $\Sigma^{0,0} = I$. For $t = 1, \cdots, T$, the first parameterization is given by $\Sigma^{t,0} = \sigma_{t-1} I$ with $\sigma_{t-1} > 0$, while the second parameterization is given by $\Sigma^{t,0} = \diag(d_{t-1})$ with $d_{t-1} \in \mathbb{R}^{n_x}$ and $d_{t-1} >0$. In the first case, $\bfSigma \tildeww$ represents the Cartesian product of $\ell_\infty$ norm balls with radii $\sigma_{t-1} > 0$; in the second case, it represents the Cartesian product of hyperrectangles whose edge lengths are given by the entries in $d_{t-1}$ times $2$. We want to search the filter parameters $\sigma_{t}$ or $d_{t}$ such that at time $t$, the reachable set of the lumped uncertainty $\eta_t$ is bounded by the simple geometric sets as shown in Fig.~\ref{fig:ball_bound}. Then, we can tighten the constraints of the robust OCP~\eqref{eq:robustOCP} efficiently using the surrogate uncertain linear dynamics~\eqref{eq:surrogate_dynamics} which only has additive disturbances.

As shown in Fig.~\ref{fig:ball_bound}, bounding the reachable set of the lumped uncertainty $\bfeta$ by the reachable set of the additive disturbances $\bfSigma \tildeww$ necessarily introduces conservatism (indicated by the gap between the blue and green regions in Fig.~\ref{fig:ball_bound}). To reduce the conservatism in uncertainty over-approximation while maintaining numerical tractability, in this work, we parameterize the filter $\bfSigma$ as follows: the sub-diagonal blocks $\Sigma^{t,t-i}$ are non-zero, and the diagonal blocks are parameterized by $\Sigma^{0,0} = I,  \Sigma^{t, 0} = \diag(d_{t-1})$ for $1 \leq t \leq T$ where $d_{t-1} \in \mathbb{R}^{n_x}$ and $d_{t-1} > 0$. We refer to this parameterization as the full parameterization of the filter. It has the following features: (a) By construction, $\bfSigma$ is invertible. (b)  It contains the aforementioned diagonal parameterization as special cases. (c) It utilizes most of the degrees of freedom offered by the block-lower-triangular parameterization of $\bfSigma$ since all off-diagonal matrices are free variables. A numerical comparison of the full and diagonal parameterizations of the filter is given in Section~\ref{sec:filter_comparison}. With the full parameterization of the filter, we are now ready to find sufficient conditions for $\mathcal{R}(\bfeta;\{\Phix, \Phiu\}) \subseteq \mathcal{R}(\bfSigma \tildeww)$.

\subsection{Reachable set over-approximation problem}
The reachable set over-approximation problem can be stated as finding $\{\Phix, \Phiu, \bfSigma\}$ such that 
\begin{equation}\label{eq:set_over_approx}
\mathcal{R}(\bfeta;\{\Phix, \Phiu\}) \subseteq \mathcal{R}(\bfSigma \tildeww)
\end{equation}
for all $(\Delta_A, \Delta_B) \in \mathcal{P}$ and $w_t \in \mathcal{W}$. The following lemma gives a rigorous formulation of constraint~\eqref{eq:set_over_approx}, and its proof is given in Appendix~\ref{app:proof_lemma}.

\begin{lemma} \label{lem:equivalence}
Let the uncertainty assumption $(\Delta_A, \Delta_B) \in \mathcal{P}$ and $w_t \in \mathcal{W}$ hold for all $0 \leq t \leq T-1$. For a filter $\bfSigma$ and LTV state feedback controller $\KK$ with induced system responses $\{\Phix, \Phiu\}$, the following two conditions are equivalent:
\begin{enumerate}
\item $\mathcal{R}(\bfeta;\{\Phix, \Phiu\}) \subseteq \mathcal{R}(\bfSigma \tildeww)$.
\item The following system of equations 
\begin{equation} \label{eq:robust_equalities}
	Z \begin{bmatrix}
		\DDelta_A & \DDelta_B 
	\end{bmatrix} \begin{bmatrix}
		\Phix \\ \Phiu
	\end{bmatrix} \bfSigma \tildeww + \ww = \bfSigma \tildeww
\end{equation}
has a solution $\tildeww^* \in \mathcal{W}_{\tildeww}$ for all possible realizations of the uncertainty parameters $\{\DDelta_A, \DDelta_B, \ww\}$.
\end{enumerate}
\end{lemma}

With the help of Lemma~\ref{lem:equivalence}, our goal is to find sufficient conditions on $\{\Phix, \Phiu\}$ and $\bfSigma$ such that~\eqref{eq:robust_equalities} is robustly feasible. However, the presence of bilinear terms $\Phix \bfSigma$ and $\Phiu \bfSigma$ in~\eqref{eq:robust_equalities} makes it challenging. To resolve this issue, we apply the change of variables
\begin{equation} \label{eq:change_var}
	\tildePhix = \Phix \bfSigma, \quad \tildePhiu = \Phiu \bfSigma 
\end{equation}
where $\bfSigma$ is invertible and $\tildePhix, \tildePhiu \in \mathcal{L}_{TV}$. In this case, guaranteeing robust feasibility of~\eqref{eq:robust_equalities} is equivalent to finding $(\tildePhix, \tildePhiu, \bfSigma)$ such that 
\begin{equation}\label{eq:robust_feasibility}
Z \begin{bmatrix}
\DDelta_A & \DDelta_B 
\end{bmatrix} \begin{bmatrix}
\tildePhix \\ \tildePhiu
\end{bmatrix}  \tildeww + \ww = \bfSigma \tildeww,\ \tildeww \in \mathcal{W}_{\tildeww}
\end{equation}
is feasible for all $(\Delta_A, \Delta_B) \in \mathcal{P}$ and $w_t\in \mathcal{W}$. It can be easily verified that all achievable $\{\tildePhix, \tildePhiu\}$ are parameterized by the following affine constraint 
\begin{equation} \label{eq:affine_scaled}
	\begin{bmatrix} I - Z \hat{\sA} & -Z \hat{\sB} \end{bmatrix} 
	\begin{bmatrix} \tildePhix\\ \tildePhiu \end{bmatrix} = \bfSigma, \ \tildePhix, \tildePhiu \in \mathcal{L}_{TV}, 
\end{equation}
which directly follows from the affine constraint~\eqref{eq:affine_constr} that parameterizes all achievable system responses $\{\Phix, \Phiu\}$ for system~\eqref{eq:system_dyn}. Note that constraint~\eqref{eq:affine_scaled} is jointly affine in $\{\tildePhix, \tildePhiu, \bfSigma\}$. The invertibility of the filter guarantees the equivalence of searching $\{\Phix, \Phiu, \bfSigma\}$ under constraint~\eqref{eq:affine_constr} and searching $\{\tildePhix, \tildePhiu, \bfSigma\}$ under constraint~\eqref{eq:affine_scaled} for $\mathcal{R}(\bfeta;\{\Phix, \Phiu\}) \subseteq \mathcal{R}(\bfSigma \tildeww)$ to hold. 

As shown in the following corollary, $\{\tildePhix, \tildePhiu \}$ can be interpreted as system responses mapping $\tildeww$ to $(\xx, \uu)$ for the system $\xx = Z\hat{\sA} \xx + Z \hat{\sB} \uu + \bfSigma \tildeww$ in closed-loop with $\uu = \KK \xx$. The proof is given in Appendix~\ref{app:proof_corollary}.

\begin{corollary} \label{coro:scaled_equivalence}
	Let $\bfSigma \in \mathcal{L}_{TV}^{T, n_x \times n_x}$ be invertible and $\KK \in \mathcal{L}_{TV}^{T, n_u \times n_x}$ be a state feedback controller. Then, for the closed-loop dynamics
	\begin{equation}\label{eq:dyn_filtered_sig}
	\xx = Z(\hat{\sA} + \hat{\sB} \KK) \xx + \bfSigma \tildeww
	\end{equation}
	over the horizon $t = 0, \cdots, T$, we have
	\begin{enumerate}
		\item The affine subspace defined by~\eqref{eq:affine_scaled} parameterizes all achievable system responses $\xx = \tildePhix \widetilde{\ww}, \uu = \tildePhiu \widetilde{\ww}$ for system~\eqref{eq:dyn_filtered_sig}.
		\item For any block-lower-triangular matrices $\{\tildePhix, \tildePhiu \}$ satisfying \eqref{eq:affine_scaled}, the controller $\KK = \tildePhiu \tildePhix^{-1}$ achieves the desired response.
	\end{enumerate}
\end{corollary}

\subsection{Convex over-approximation constraints}
We now present sufficient conditions on $\{\tildePhix, \tildePhiu, \bfSigma\}$ such that the robust feasibility of~\eqref{eq:robust_feasibility} is guaranteed. The block-downshift operator $Z$ makes it possible to decompose and analyze the equality constraints in~\eqref{eq:robust_feasibility} for $t = 0, \cdots, T-1$ sequentially. 

\subsubsection{Case $t=0$}
For bounding $\eta_0$ at $t = 0$, we have the following constraint from~\eqref{eq:robust_feasibility}
\begin{equation} \label{eq:constr_0}
\Delta_A x_0 + \Delta_B \tilde{\Phi}_u^{0,0} x_0 + w_0 = \Sigma^{1,1} x_0 + \Sigma^{1,0} \tilde{w}_0
\end{equation}
where we have used $\tilde{\Phi}_x^{0,0} = I$ as a result of~\eqref{eq:affine_scaled} and $\Sigma^{0,0} = I$. The above constraint is robustly feasible with a solution $\lVert \tilde{w}_0 \rVert_\infty \leq 1$ if and only if 
\begin{equation} \label{eq:constr_0_transformed}
\lVert {\Sigma^{1,0}}^{-1} (\Delta_A x_0 + \Delta_B \tilde{\Phi}_u^{0,0} x_0 - \Sigma^{1,1} x_0 + w_0) \Vert_\infty \leq 1
\end{equation}
for all $(\Delta_A, \Delta_B) \in \mathcal{P}$ and $w_0 \in \mathcal{W}$. Constraint~\eqref{eq:constr_0_transformed} is non-convex in $\Sigma^{1,0}$, but with our diagonal matrix parameterization $\Sigma^{1,0} = \diag(d_0)$ (see Section~\ref{sec:filter_parameterization}) it can be rewritten as 
\begin{equation} \label{eq:constr_0_convex}
\lvert e_i^\top(\Delta_A x_0 + \Delta_B \tilde{\Phi}_u^{0,0} x_0 - \Sigma^{1,1} x_0 + w_0) \rvert \leq d_{0, i}
\end{equation}
for $i = 1, \cdots, n_x$ where $d_{0,i}$ denotes the $i$-th entry of $d_0$ and $e_i$ is the $i$-th {standard basis}. We compute offline~\footnote{The parameters $\sigma_{w,i}$ for $i = 1, \dots, n_x$ only need to be computed once by solving $2 n_x$ simple linear programs. Afterward they are used for solving the robust OCP~\eqref{eq:robustOCP} for all time steps.}
\begin{equation} \label{eq:offline_sigma}
\sigma_{w,i} = \max\{ h_\mathcal{W}(e_i), h_\mathcal{W}(-e_i)\}, i = 1, \cdots, n_x
\end{equation}
where $h_\mathcal{W}(\cdot)$ is the support function of $\mathcal{W}$.
Then, the following constraints
\begin{equation}\label{eq:constr_0_convex_design}
\begin{aligned}
&\lvert e_i^\top (\Delta_A x_0 + \Delta_B \tilde{\Phi}_u^{0,0} x_0 - \Sigma^{1,1} x_0 ) \rvert + \sigma_{w,i} \leq d_{0,i}, \quad \forall (\Delta_A, \Delta_B) \in \vertex(\mathcal{P}), \quad i = 1, \cdots, n_x,
\end{aligned}
\end{equation}
guarantee that~\eqref{eq:constr_0_convex} is robustly feasible, where $\vertex(\mathcal{P}) := \{ (\Delta_{A,j}, \Delta_{B,j}), j = 1, \cdots, M\}$ denotes the set of vertices of $\mathcal{P}$. To prove this, we can show that the left-hand-side (LHS) of~\eqref{eq:constr_0_convex_design} is an upper bound on the LHS of~\eqref{eq:constr_0_convex} by using the triangle inequality of the absolute value $\lvert \cdot \rvert$, the definition of $\sigma_{w,i}$ in~\eqref{eq:offline_sigma}, and the fact that the LHS of~\eqref{eq:constr_0_convex} is convex in $(\Delta_A, \Delta_B)$ and the maximum of a convex function over a convex polytope is achieved at the vertices~\citep{boyd2004convex}. 

Constraint~\eqref{eq:constr_0_convex_design} is convex in the design parameters $\tilde{\Phi}_u^{0,0}, \Sigma^{1,1}$ and $d_1$. It guarantees that for all possible realizations of $(\Delta_A, \Delta_B, w_0)$ and the generated lumped uncertainty $\eta_0$, we can always find $\tilde{w}_0^*$ such that $\eta_0 = \Sigma^{1,1} x_0 + \Sigma^{1,0} \tilde{w}_0^*$ with $\lVert \tilde{w}_0^* \rVert_\infty \leq 1$. We fix $\tilde{w}_0^*$ as the $\tilde{w}_0$-component of the solution $\tildeww^*$ to~\eqref{eq:robust_feasibility}.

\subsubsection{Case: $t = 1$} 
Similarly, to bound $\eta_1$, we can write the relevant equality constraints from~\eqref{eq:robust_feasibility} as
\begin{equation}\label{eq:constr_1} 
\begin{aligned}
\Delta_A(\tilde{\Phi}_x^{1,1} x_0 + \tilde{\Phi}_x^{1,0} \tilde{w}_0^*) + \Delta_B(\tilde{\Phi}_u^{1,1} x_0 + \tilde{\Phi}_u^{1,0} \tilde{w}_0^*) + w_1 = \Sigma^{2,2} x_0 + \Sigma^{2,1} \tilde{w}_0^* + \Sigma^{2,0} \tilde{w}_1
\end{aligned}
\end{equation}
where $\tilde{w}_0^*$ is the solution from the previous time step and captures the effects of $\eta_0$ on future perturbations $\eta_t$ for $t \geq 1$. Since $\Sigma^{2,0} = \diag(d_1)$ is a diagonal matrix, following the same steps as in the case $t = 0$ and grouping the terms in~\eqref{eq:constr_1} by $x_0, \tilde{w}_0^*,$ and $\tilde{w}_1$, we conclude if the inequalities
\begin{equation}\label{eq:constr_1_w}
	\begin{aligned}
	\lvert e_i^\top(\Delta_A\tilde{\Phi}_x^{1,1}  + \Delta_B \tilde{\Phi}_u^{1,1} - \Sigma^{2,2})x_0 \rvert  + \lvert e_i^\top(\Delta_A\tilde{\Phi}_x^{1,0}  + \Delta_B \tilde{\Phi}_u^{1,0} - \Sigma^{2,1})\tilde{w}_0^* \rvert + \lvert e_i^\top w_1 \rvert \leq d_{1,i}
	\end{aligned}
\end{equation}
for $1 \leq i \leq n_x$ hold robustly, then the robust feasibility of~\eqref{eq:constr_1} is guaranteed. However, the exact value of $\tilde{w}_0^*$ is unknown to us since it depends on $(\Delta_A, \Delta_B, w_0)$. To address this issue, we treat $\tilde{w}_0^*$ as uncertainty satisfying $\lVert \tilde{w}_0^* \rVert_\infty \leq 1$ and further tighten the constraint~\eqref{eq:constr_1_w} as
\begin{equation}\label{eq:constr_1_convex_design}
\begin{aligned}
& \lvert e_i^\top(\Delta_A\tilde{\Phi}_x^{1,1}  + \Delta_B \tilde{\Phi}_u^{1,1} - \Sigma^{2,2})x_0 \rvert  + \lVert e_i^\top(\Delta_A\tilde{\Phi}_x^{1,0}  + \Delta_B \tilde{\Phi}_u^{1,0} - \Sigma^{2,1})\rVert_1 + \sigma_{w,i} \leq d_{1,i}, \\
& \forall (\Delta_A, \Delta_B) \in \vertex(\mathcal{P}), \quad i = 1, \cdots, n_x,
\end{aligned}
\end{equation}
by applying the H\"older's inequality $\lvert  a^\top \tilde{w}_0^* \rvert \leq \lVert a \rVert_1 \lVert \tilde{w}_0^* \rVert_\infty \leq \lVert a \rVert_1$ and using the fact that $\lvert e_i^\top w_1 \rvert \leq \sigma_{w,i}$. 


\subsubsection{General case}
We repeat this process from $t = 0$ to $t = T-1$ to obtain a set of convex constraints on $\{\tildePhix, \tildePhiu, \bfSigma\}$: 
\begin{equation}\label{eq:over_approx_constr}
	\begin{aligned}
	&  \lvert e_i^\top (\Delta_A \tilde{\Phi}_x^{t,t} + \Delta_B \tilde{\Phi}_u^{t,t}- \Sigma^{t+1,t+1})x_0 \rvert +  \sigma_{w,i} + \sum_{i=1}^t \lVert e_i^\top (\Delta_A \tilde{\Phi}_x^{t,t-i} + \Delta_B \tilde{\Phi}_u^{t,t-i}- \Sigma^{t+1,t+1-i})\rVert_1  \leq d_{t, i}, \\
	& \forall (\Delta_A, \Delta_B) \in \vertex(\mathcal{P}), i = 1, \cdots, n_x, t = 0, \cdots, T-1.
	\end{aligned}
\end{equation}
In fact, the constraints in~\eqref{eq:over_approx_constr} can be translated into an equivalent set of linear constraints on $(\tildePhix, \tildePhiu, \bfSigma)$. 

By the derivation of~\eqref{eq:over_approx_constr}, any feasible solution $(\tildePhix, \tildePhiu, \bfSigma)$ to~\eqref{eq:over_approx_constr} guarantees the robust feasibility of~\eqref{eq:robust_feasibility}. By Lemma~\ref{lem:equivalence}, they also guarantee that $\mathcal{R}(\bfeta;\{\Phix, \Phiu\}) \subseteq \mathcal{R}(\bfSigma \tildeww)$. In this case, it suffices to consider the uncertain dynamical system~\eqref{eq:surrogate_dynamics} with the surrogate additive disturbance $\bfSigma \tildeww$ for solving the robust OCP~\eqref{eq:robustOCP}. The synthesized  controller $\KK = \tildePhiu \tildePhix^{-1} = \Phiu \Phix^{-1}$ guarantees robust constraint satisfaction for the original uncertain dynamical system~\eqref{eq:system_dyn}. We summarize all the steps of solving the robust OCP in the next section.


\section{Formulation of SLS MPC}
\label{sec:convex_OCP}
In this section, we present our solution to the robust OCP~\eqref{eq:robustOCP} and discuss the closed-loop properties of the proposed robust MPC method. 

\subsection{Constraint tightening of robust OCP}
Recall that under constraints~\eqref{eq:affine_scaled} and~\eqref{eq:over_approx_constr}, it suffices to consider the uncertain dynamics $\xx = Z\hat{\sA} \xx + Z \hat{\sB} \uu + \bfSigma \tildeww$ with $\lVert \tilde{w}_t \rVert_\infty \leq 1$ for guaranteeing robust constraint satisfaction of the synthesized controller $\KK = \tildePhiu \tildePhix^{-1}$ since the filtered disturbance signal $\bfSigma \tildeww$ can realize all possible values of the lumped uncertainty $\bfeta$. 

Let us take the state constraint tightening of $x_t \in \mathcal{X}$ as an example. The state constraint is a polyhedral set $\mathcal{X} = \lbrace x \in \mathbb{R}^{n_x} \mid F_x x \leq b_x \rbrace$. Denote the number of linear constraints in defining $\mathcal{X}$ as $n_\mathcal{X}$ and $\facet(\mathcal{X}) = \{(F_x(:,i), b_x(i)) \mid i = 1, \cdots, n_\mathcal{X} \}$ as the set of all linear constraint parameters of $\mathcal{X}$. For the dynamical system~\eqref{eq:surrogate_dynamics} with filtered disturbance $\bfSigma \tildeww$, the affine constraint~\eqref{eq:affine_scaled} parameterizes all achievable closed-loop system responses $(\tildePhix,\tildePhiu)$ such that $\xx = \tildePhix \tildeww , \uu = \tildePhiu \tildeww$ under an LTV state feedback controller $\uu = \KK \xx$ (see Corollary~\ref{coro:scaled_equivalence}). Using $x_t = \tilde{\Phi}_x^{t,t}x_0 + \sum_{i=1}^t \tilde{\Phi}_x^{t,t-i} \tilde{w}_{i-1}$, the following tightened state constraints
\begin{equation} \label{eq:state_constr_tightening}
\begin{aligned}
f^\top \widetilde{\Phi}_x^{t,t} x_0 + \sum_{i=1}^t \lVert f^\top \widetilde{\Phi}_x^{t,t-i} \rVert_1 \leq b, \forall (f, b) \in \text{facet}(\mathcal{X}), \ t = 0, \cdots, T-1
\end{aligned}
\end{equation}
guarantee $f^\top x_t \leq b$ for all $(f, b) \in \text{facet}(\mathcal{X})$ robustly under the controller $\KK = \tildePhiu \tildePhix^{-1}$. This follows from a direct application of the H\"older's inequality on $f^\top x_t$ and the fact $\lVert \tilde{w}_t \rVert_\infty \leq 1$. Similarly, we can tighten the terminal constraint as
\begin{equation} \label{eq:terminal_constr_tightening}
f^\top \widetilde{\Phi}_x^{T,T} x_0 + \sum_{i=1}^T \lVert f^\top \widetilde{\Phi}_x^{T,T-i} \rVert_1 \leq b, \ \forall (f, b) \in \text{facet}(\mathcal{X}_T),
\end{equation}
and tighten the control input constraints as
\begin{equation}\label{eq:input_constr_tightening}
\begin{aligned}
f^\top \widetilde{\Phi}_u^{t,t} x_0 + \sum_{i=1}^t \lVert f^\top \widetilde{\Phi}_u^{t,t-i} \rVert_1 \leq b, \ \forall (f, b) \in \text{facet}(\mathcal{U}), \ t = 0, \cdots, T-1.
\end{aligned}
\end{equation}
Again, constraints~\eqref{eq:state_constr_tightening}, \eqref{eq:terminal_constr_tightening}, \eqref{eq:input_constr_tightening} can be easily expressed as linear constraints on $\{\tildePhix, \tildePhiu\}$.

\subsection{Convex tightening of the robust OCP}
\label{sec:convex_OCP_formulation}
Summarizing all the aforementioned steps, we can synthesize a robust LTV state feedback controller $\KK$ as shown in the following theorem. 
{
\begin{theorem} \label{thm:convex}
Consider the convex quadratic program
\begin{equation} \label{eq:convex_inner_approx}
\begin{aligned}
\tilde{J}^*_T = &\underset{\tildePhix, \tildePhiu, \bfSigma}{\textrm{min}}  \quad \tilde{J}_T(\tildePhix, \tildePhiu) \\
\textrm{s.t.} & \quad \textrm{affine constraint~\eqref{eq:affine_scaled}} \\
& \quad \textrm{over-approximation constraint~\eqref{eq:over_approx_constr}} \\
& \quad \textrm{tightened constraints~\eqref{eq:state_constr_tightening}, \eqref{eq:terminal_constr_tightening}, \eqref{eq:input_constr_tightening}} \\
&\quad x_0 = x(k)
\end{aligned}
\end{equation}
where $\tildePhix \in \mathcal{L}_{TV}^{T, n_x \times n_x}$, $\tildePhiu \in \mathcal{L}_{TV}^{T, n_u \times n_x}$, $\bfSigma \in \mathcal{L}_{TV}^{T, n_x \times n_x}$ is parameterized in Section~\ref{sec:filter_parameterization}, and the cost function is given by
\begin{equation} \label{eq:tilde_cost}
\tilde{J}_T(\tildePhix, \tildePhiu) = 	 \Big \lVert \begin{bmatrix}
		\mathbf{Q}^{1/2} & \\ & \mathbf{R}^{1/2}
	\end{bmatrix} \begin{bmatrix}
		\tildePhix(:,0) \\ \tildePhiu(:,0)
	\end{bmatrix}x_0 \Big \rVert_2^2 
\end{equation}
with $\mathbf{Q} = \blkdiag(Q, \cdots, Q, Q_T)$, $\mathbf{R} =\blkdiag(R, \cdots, R, 0)$. We have:
\begin{enumerate}
\item For any feasible solution $\{\tildePhix, \tildePhiu, \bfSigma\}$ of Problem~\eqref{eq:convex_inner_approx}, the LTV state feedback controller $\KK= \tildePhiu \tildePhix^{-1}$ is feasible for the robust OCP~\eqref{eq:robustOCP}.
\item With the additional structural constraint $\Sigma^{t,t} = 0, t = 1, \cdots, T$ on the filter $\bfSigma$, the optimal cost in~\eqref{eq:convex_inner_approx} is an upper bound on the optimal cost of the robust OCP~\eqref{eq:robustOCP}.
\end{enumerate}
\end{theorem}

\textbf{Proof}: The proof of the first point follows directly from our derivation of the constraints in~\eqref{eq:convex_inner_approx} from the previous sections. To prove the second point, denote $J_T^*$ the optimal cost of Problem~\eqref{eq:robustOCP} and $\tilde{J}_T^*$ the optimal cost of Problem~\eqref{eq:convex_inner_approx}. With the constraint $\Sigma^{t,t} = 0$, $t= 1, \cdots, T$, the nominal dynamics of the surrogate system~\eqref{eq:surrogate_dynamics}, i.e., when $\tilde{w}_t = 0$ for $t = 0, \cdots, T-1$, equals that of the actual uncertain system~\eqref{eq:system_dyn}. Since the set of feasible solutions ${\tildePhix, \tildePhiu}$ to Problem~\eqref{eq:convex_inner_approx} only constitutes a subset of all robust LTV state feedback controllers for the robust OCP~\eqref{eq:robustOCP}, we have $\tilde{J}_T^* \geq J_T^*$. 

Note that since each entry of $d_t$ is lower bounded by $\sigma_w >0$ from~\eqref{eq:over_approx_constr}, any feasible $\bfSigma$ of Problem~\eqref{eq:convex_inner_approx} is invertible. Theorem~\ref{thm:convex} shows that by solving the quadratic program~\eqref{eq:convex_inner_approx}, we can search for robust LTV state feedback controllers while minimizing an upper bound on $J_T^*$. Using the surrogate dynamics~\eqref{eq:surrogate_dynamics}, we can further extend our framework to minimize an upper bound on the worst-case costs of the robust OCP~\eqref{eq:robustOCP}.

\begin{corollary} \label{coro:min-max}
Consider the worst-case cost in the robust OCP~\eqref{eq:robustOCP} :
\begin{equation} \label{eq:min-max}
	\begin{aligned}
	& J^*_{T,w} = \min_{\pi_{0:T-1}} \max_{ \substack{(\Delta_A, \Delta_B) \in \mathcal{P} \\ w_{t} \in \mathcal{W}, 0 \leq t \leq T-1 } } \Big \lVert \begin{bmatrix}
	\mathbf{Q}^{1/2} & \\ & \mathbf{R}^{1/2}
\end{bmatrix} \begin{bmatrix}
\xx \\ \uu
\end{bmatrix} \Big \rVert_\infty.
	\end{aligned}
\end{equation}
Let $\tilde{J}_{T,w}^*$ be the optimal cost of Problem~\eqref{eq:convex_inner_approx} with 
\begin{equation} \label{eq:surrogate_min_max}
	\begin{aligned}
	\tilde{J}_T(\tildePhix, \tildePhiu) =& \Big \lVert \begin{bmatrix}
		\mathbf{Q}^{1/2} & \\ & \mathbf{R}^{1/2}
	\end{bmatrix} \begin{bmatrix}
		\tildePhix(:,0) \\ \tildePhiu(:,0)
	\end{bmatrix}x_0 \Big \rVert_\infty + \\
&    \Big \lVert \begin{bmatrix}
		\mathbf{Q}^{1/2} & \\ & \mathbf{R}^{1/2}
	\end{bmatrix} \begin{bmatrix}
		\tildePhix(:,1:T) \\ \tildePhiu(:,1:T)
	\end{bmatrix} \Big \rVert_\infty.
	\end{aligned}
\end{equation}
Then, we have $\tilde{J}_{T,w}^* \geq J_{T, w}^*$. 
\end{corollary}

\textbf{Proof}: Through the triangle inequality and the multiplicativity of the $\ell_\infty$ norm, we have that $\tilde{J}_T(\tildePhix, \tildePhiu)$ defined in~\eqref{eq:surrogate_min_max} is an upper bound on 
\begin{equation} \label{eq:inf_upper_bound}
	\max_{ \substack{\lVert \tilde{w}_t\rVert_\infty \leq 1, \\ 0 \leq t \leq T-1}} \Big \lVert \begin{bmatrix}
		\mathbf{Q}^{1/2} & \\ & \mathbf{R}^{1/2}
	\end{bmatrix} \begin{bmatrix}
		\tildePhix \\ \tildePhiu
	\end{bmatrix} \tildeww \Big \rVert_\infty,
\end{equation}
which is the worst-case cost for the surrogate system~\eqref{eq:surrogate_dynamics} under the controller $\KK = \tildePhiu \tildePhix^{-1}$. For any feasible solution $\{\tildePhix, \tildePhiu, \bfSigma\}$ to Problem~\eqref{eq:convex_inner_approx}, the filtered disturbance $\bfSigma \tildeww$ in the surrogate dynamics~\eqref{eq:surrogate_dynamics} over-approximates the lumped uncertainty in the actual uncertain system~\eqref{eq:system_dyn}. Therefore, Eq.~\eqref{eq:inf_upper_bound} is an upper bound on $J_{T,w}^*$ and we have $\tilde{J}_T(\tildePhix, \tildePhiu) \geq J_{T,w}^*$. Taking the minimum over all feasible $\{\tildePhix, \tildePhiu, \bfSigma\}$ gives $\tilde{J}_{T,w}^* \geq J_{T,w}^*$.

\begin{remark}
When the $\ell_2$ norm is used to define the worst-case cost $J_{T,w}^*$ in~\eqref{eq:min-max}, we can assume the virtual disturbances $\tilde{w}_t$ are bounded in $\ell_2$ norm rather than $\ell_\infty$ norm to simplify the derivation of an upper bound on $J_{T,w}^*$. With $\lVert \tilde{w}_t \rVert_2 \leq 1$, the derivation of constraints~\eqref{eq:over_approx_constr} to~\eqref{eq:input_constr_tightening} follows similarly as in the $\ell_\infty$ case~\footnote{With $\lVert \tilde{w}_t \rVert_2 \leq 1$, the diagonal blocks of the filter $\bfSigma$ now can only be parameterized as $\Sigma^{t, 0} = \sigma_{t-1} I$ rather than $\Sigma^{t, 0} = \diag(d_{t-1})$.  This allows rewriting constraints like~\eqref{eq:constr_0_transformed} as convex ones even when the $\ell_2$ norm is used.}.  Then, define $\tilde{J}_T(\tildePhix, \tildePhiu)$ as in~\eqref{eq:surrogate_min_max} but with the $\ell_\infty$ and $\ell_\infty$-induced norms replaced by their $\ell_2$ counterparts, and let $\tilde{J}_{T,w}^*$ denote the corresponding optimal cost of Problem~\eqref{eq:convex_inner_approx}. Following the proof of Corollary~\ref{coro:min-max}, we have $\tilde{J}_{T,w}^*  \geq J^*_{T,w}$. 
\end{remark}

\subsection{Closed-loop properties}
\label{sec:cl_properties}
SLS MPC solves the OCP~\eqref{eq:convex_inner_approx} at each time instant and applies the first optimal control input to drive the system~\eqref{eq:dyn} to the next state.  However, as in the uncertainty over-approximation-based methods~\citep{bujarbaruah2021simple, bujarbaruah2022robust}, it is challenging to show that the OCP~\eqref{eq:convex_inner_approx} is recursively feasible with a fixed horizon $T$. To illustrate this, let $\eta_{t \mid k}$ denote the predicted lumped uncertainty at time $t$ in the robust OCP with the initial state $x(k)$. We note that the constraints in~\eqref{eq:convex_inner_approx} are time-varying, which means the constraints on bounding $\eta_{t \mid k}$ and $\eta_{t \mid k+1}$ are different, defying the use of the standard shifting argument~\citep[Chapter 12]{borrelli2017predictive} in the space of system responses to prove recursive feasibility. On the other hand, the use of the conservative uncertainty over-approximation~\eqref{eq:over_approx_constr} implies that the search space of~\eqref{eq:convex_inner_approx} does not include all robustly feasible LTV state feedback controllers $\uu = \KK\xx$. Therefore, the shifting argument in the controller space cannot be used either. 

To provide closed-loop guarantee of constraint satisfaction and Input-to-State Stability (ISS), we can equip SLS MPC with the shrinking horizon strategy shown in~\citet{bujarbaruah2022robust}. In short, at time $k=0$, we solve Problem~\eqref{eq:convex_inner_approx} with horizon $T = N$ and a robust forward invariant set $\mathcal{X}_T$ as the terminal set. We can use the feasible solution at $k=0$ to construct a safe backup policy $\uu = \KK \xx$ which guarantees that we can drive the system state into $\mathcal{X}_T$ in at most $N$ steps.  {Specifically, at time $0 < k < N$, we can apply the controller $u(k) = \sum_{i=0}^{k} K^{k, k-i} x(i) $ where $K^{k,k-i}$ are the corresponding matrices drawn from backup policy $\KK$.} Then, at time $k  = N, N+1, \cdots$, the robust OCP~\eqref{eq:robustOCP} rather than Problem~\eqref{eq:convex_inner_approx} is solved exactly with horizon $T = 1$, i.e., no conservative uncertainty over-approximation is used, by enumerating the vertices of the model uncertainty set. Due to the robust forward invariance of $\mathcal{X}_T$, the robust OCP~\eqref{eq:robustOCP} with horizon $T = 1$ is always feasible and the closed-loop trajectory will remain inside $\mathcal{X}_T$ while observing all the state and input constraints for $k > N$. We refer the readers to~\citet{bujarbaruah2022robust} for the details of this strategy and the proof of robust constraint satisfaction and ISS of the closed-loop system. 
}

\section{Numerical Comparison}
\label{sec:simulation}
In this section, we compare the proposed SLS MPC with existing robust MPC methods in terms of conservatism (measured by the size of the feasible domain) and computational complexity (measured by solver time) through extensive simulation. Two classes of baselines, i.e., the tube-based~\citep{langson2004robust, lorenzen2019robust, kohler2019linear, lu2019robust} and uncertainty over-approximation-based methods~\citep{bujarbaruah2021simple, bujarbaruah2022robust}, are considered. The main features of each method are summarized in Table~\ref{table:tube} and Table~\ref{table:state_feedback}. Through the numerical examples, we demonstrate that 
\begin{enumerate}
\item \textbf{Conservatism} (measured by the size of the feasible domain): SLS MPC consistently outperforms all baseline methods and by a large margin in the face of large uncertainty. Under varying uncertainty parameters, the feasible domain of SLS MPC is always more than $90\%$ of the maximal robust control invariant set (the theoretically largest feasible domain for any robust MPC method) in the tested examples while the baseline methods become overly conservative quickly as the uncertainty becomes large.
 
\item \textbf{Computational complexity}: SLS MPC achieves comparable solver time as the baselines. Specifically, it has the same level of complexity as the uncertainty over-approximation-based methods.
\end{enumerate}

\begin{table*}[ht]
	\centering
	\caption{Tube-based MPC methods.}
	\label{table:tube}
	\begin{tabular}{*{4}{c}}
		\toprule
		Method & Controller & Tube & Vertex enumeration of tube \\ \midrule
		\texttt{Tube-A}~\citep{langson2004robust} & Barycentric & Homothetic & Yes   \\ \midrule
		\texttt{Tube-B}~\citep{lorenzen2019robust} & $Kx_t + v_t$ & Homothetic & Yes \\ \midrule
		\texttt{Tube-C}~\citep{kohler2019linear} & $Kx_t + v_t$ & Homothetic & No  \\ \midrule 
		\texttt{Tube-D}~\citep{lu2019robust} &$Kx_t + v_t$ & Hyperplane & No   \\
		\bottomrule
	\end{tabular}
\end{table*}

\begin{table*}[ht]
	\centering
	\caption{Uncertainty over-approximation-based MPC methods.}
	\label{table:state_feedback}
	\begin{tabularx}{\textwidth}{X X}
		\toprule
		Method & Features \\ \midrule
		\texttt{Lumped-Disturbance-MPC}~\citep{bujarbaruah2021simple} & Over-approximate lumped uncertainty globally by a norm-bounded additive disturbance signal.  \\ \midrule
		\texttt{Offline-Tightening-MPC}~\citep{bujarbaruah2022robust} & Compute constraint tightening margins offline. Search controller online. \\ \midrule
		\texttt{SLS-MPC} (our method) & Search lumped uncertainty over-approximation and controller jointly in the space of system responses.  \\ 
		\bottomrule
	\end{tabularx}
\end{table*}

\subsection{Interpretation of the results}
SLS MPC achieves significant improvement in conservatism thanks to (a) the LTV  state feedback controller parameterization $\uu = \KK \xx$, and (b) the novel uncertainty over-approximation method (see Section~\ref{sec:over_approx}). We note that there is naturally a tension between controller parameterization and constraint tightening in solving the robust OCP~\eqref{eq:robustOCP}, since considering a more complex controller class necessarily makes it harder to guarantee robust satisfaction of constraints. The performance of SLS MPC is attributed to balancing both aspects in a desirable manner.

\textbf{Conservatism}: As shown in Table~\ref{table:tube}, existing tube-based methods are mainly restricted to using a relatively simple controller parameterization $u_t = K x_t + v_t$ where the time-invariant pre-stabilizing feedback gain $K$ is chosen offline instead of optimized online. 
In contrast, SLS MPC utilizes the LTV state feedback controller $\uu = \KK \xx$ which contains $u_t = K x_t + v_t$ as a subclass and searches for $\KK$ online. Compared with the uncertainty over-approximation-based methods~\citep{bujarbaruah2021simple, bujarbaruah2022robust} shown in Table~\ref{table:state_feedback}, SLS MPC adopts a novel constraint tightening approach that simultaneously over-approximates the lumped uncertainty by a virtual filtered additive disturbance signal and searches for the controller. 

\textbf{Computational complexity}: The use of LTV state feedback controllers necessarily leads to the quadratic growth of the number of decision variables in the horizon. This holds true for the uncertainty over-approximation-based methods~\citep{bujarbaruah2021simple, bujarbaruah2022robust} and SLS MPC. In contrast, the number of decision variables is often linear in the horizon in tube-based MPC. Therefore, we expect the solver time of the uncertainty over-approximation-based methods, including SLS MPC, to be larger than that of tube-based MPC. Fig.~\ref{fig:solver_time_comparison} supports our analysis, but it also suggests that SLS MPC may even be preferable to some tube-based methods when the uncertainty is large. 

We provide a detailed explanation of both tube-based and uncertainty over-approximation-based baseline methods and their implementation in Appendix~\ref{sec:mpc_baselines}. We note that to the best of our knowledge, such a comprehensive numerical comparison and evaluation of existing robust MPC methods have not been done before. We make our codes for implementing SLS MPC and all listed baselines publicly available at \url{https://github.com/ShaoruChen/Polytopic-SLSMPC}. All the experiments in this section were implemented in MATLAB R2019b with YALMIP~\citep{lofberg2004yalmip} and MOSEK~\citep{mosek} on an Intel i7-6700K CPU.

\subsection{Test example}
\label{sec:test_example}
We evaluate the conservatism of SLS MPC and other baseline methods on a 2-dimensional system adapted from~\citet{bujarbaruah2022robust}  in Section~\ref{sec:feasible_domain_comparison}.
The only difference from the example in~\citet{bujarbaruah2022robust} is that we use a different model uncertainty assumption, i.e., we only consider uncertainty in one entry of the dynamics matrices $A$ and $B$, to allow varying the uncertainty levels in a wide range and highlight the differences between the tested methods. {In Section~\ref{sec:random_example}, randomly generated systems are used for conservatism comparison. Experiments on the original example in~\citet{bujarbaruah2022robust} are shown in Appendix~\ref{sec:original_example}.}

The system nominal dynamics and problem constraints are given as 
\begin{equation} \label{eq:example}
	\begin{aligned}
		\hat{A} = \begin{bmatrix}
			1 & 0.15 \\ 0.1 & 1 
		\end{bmatrix}, \quad \hat{B} = \begin{bmatrix}
			0.1 \\ 1.1
		\end{bmatrix}, \quad \mathcal{X} = \Big \{ x \in \mathbb{R}^2 \vert \begin{bmatrix}
			-8 \\ -8
		\end{bmatrix} \leq x \leq \begin{bmatrix}
			8 \\8
		\end{bmatrix}  \Big \}, \quad \mathcal{U} = \{ u \in \mathbb{R} \vert -4 \leq u \leq 4\}.
	\end{aligned}
\end{equation}
We consider the following polytopic model uncertainty:
\begin{equation} \label{eq:uncertainty_form}
	\begin{aligned}
		\Delta_A \in \conv \Big \{ \begin{bmatrix}
			\epsilon_A & 0 \\ 0 & 0
		\end{bmatrix} ,  \begin{bmatrix}
			-\epsilon_A & 0 \\ 0 & 0
		\end{bmatrix} \Big \}, \quad \Delta_B \in \conv\Big \{ \begin{bmatrix}
			0 \\ \epsilon_B
		\end{bmatrix},  \begin{bmatrix}
			0 \\ -\epsilon_B
		\end{bmatrix} \Big \}
	\end{aligned}
\end{equation}
and the norm bounded additive disturbances $\lVert w_t \rVert_\infty \leq \sigma_w$ for robust MPC~\footnote{In this case, according to~\eqref{eq:offline_sigma}, we have $\sigma_{w, i}  = \sigma_w$ for $i = 1, \cdots, nx$. }. The uncertainty parameters $\epsilon_A, \epsilon_B$ and $\sigma_w$ are to be specified. The cost weights are chosen as $Q = 10I, R = 1, Q_T = 10I$. 

\subsection{Conservatism comparison}
\label{sec:simulation_conservatism}
\subsubsection{Feasible domain comparison}
\label{sec:feasible_domain_comparison}
\begin{figure}
	\centering
	\begin{subfigure}{0.49 \columnwidth}
		\includegraphics[width = \textwidth]{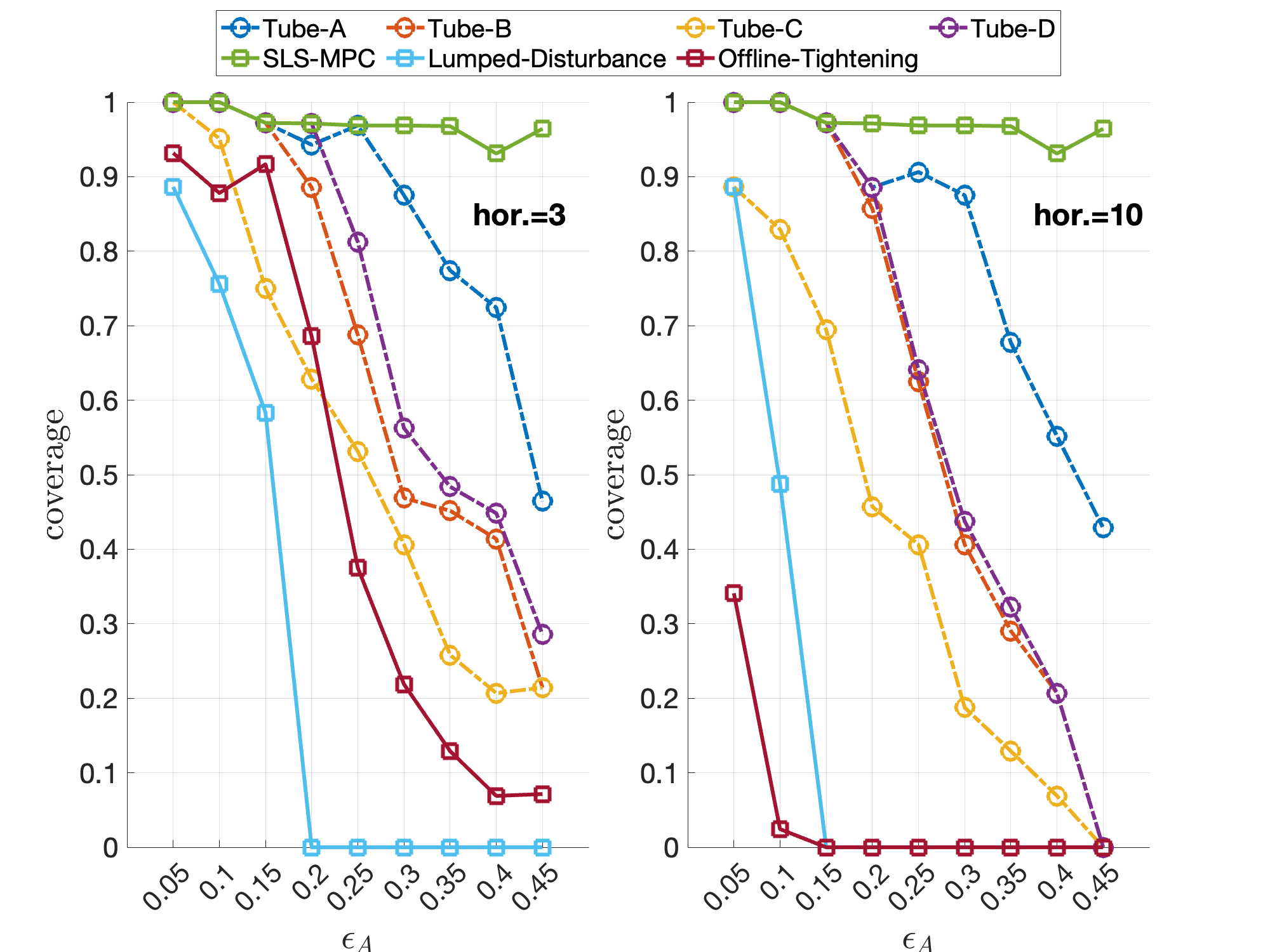}
		\caption{$\epsilon_B = 0.1, \sigma_w = 0.1$}
	\end{subfigure}
	\hfill
	\begin{subfigure}{ 0.49\columnwidth}
		\includegraphics[width = \textwidth]{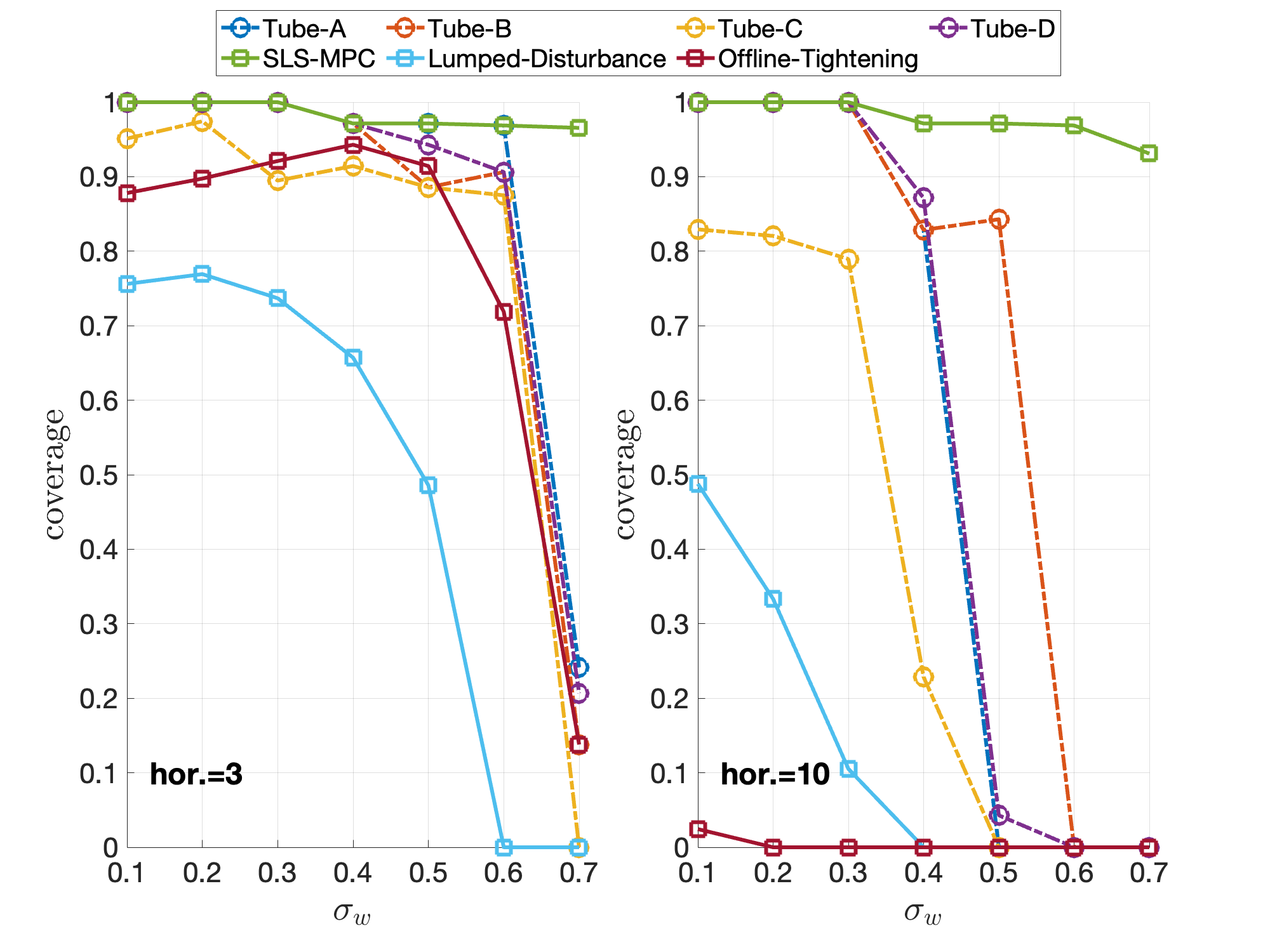}
		\caption{ $\epsilon_A = 0.1, \epsilon_B = 0.1$.}
	\end{subfigure}
	\caption{{Coverage comparison of the robust MPC methods with different uncertainty parameters (Left: with varying model uncertainty parameter $\epsilon_A$. Right: with varying disturbance bounds $\sigma_w$) and horizons $T=3$ and $T=10$. \texttt{SLS-MPC} consistently outperforms all other baselines and achieves coverages over $90\%$.}}
	\label{fig:coverage_comparison}
\end{figure}

\begin{figure}[tb]
	\centering
	\includegraphics[width = 0.6 \columnwidth]{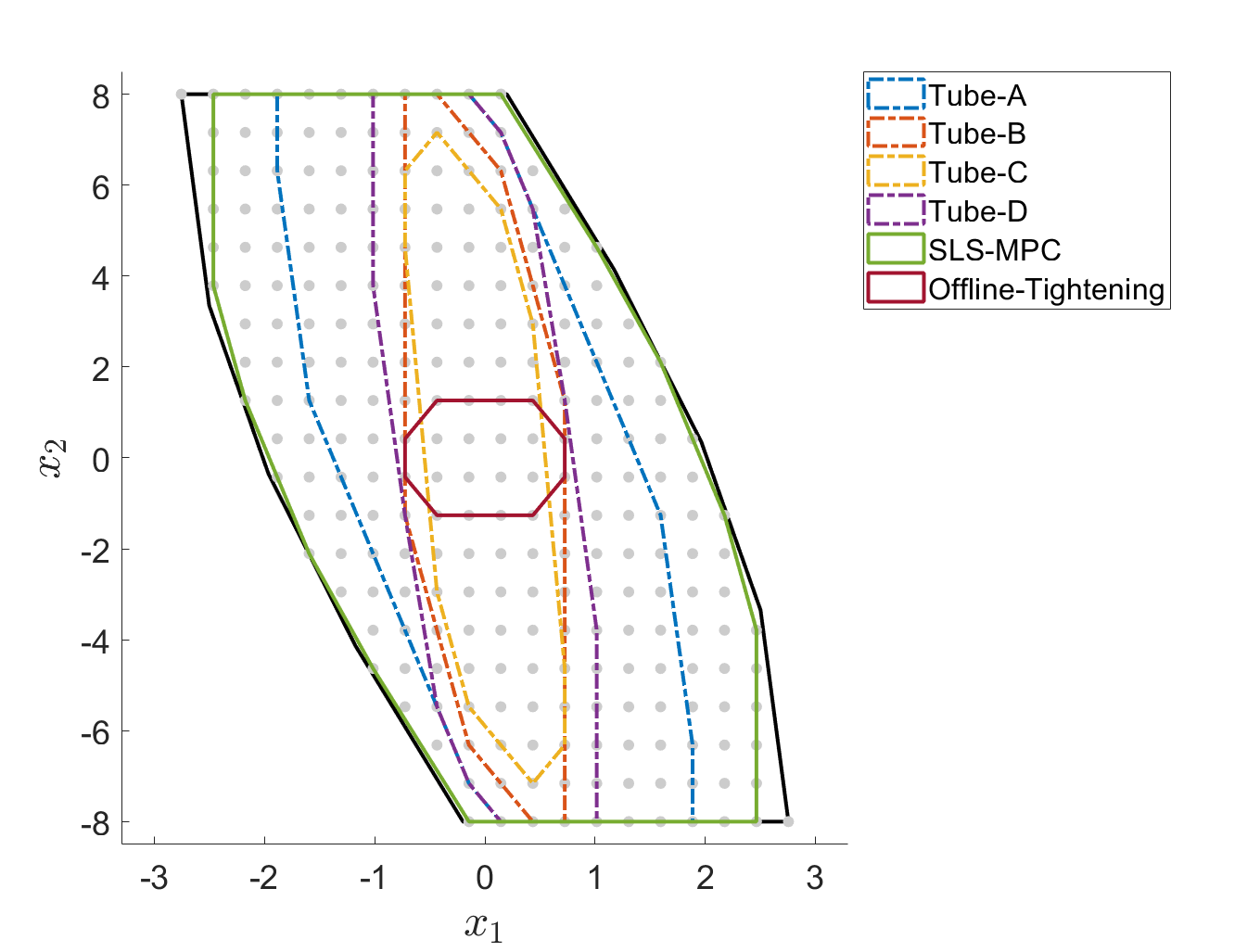}
	\caption{{Feasible domain comparison of robust MPC methods with uncertainty parameters $\epsilon_A = 0.4, \epsilon_B = 0.1, \sigma_w = 0.1$, and horizon $ T=3$. The gray dots denote the initial states sampled from a $20 \times 20$ grid \protect \footnotemark.}}
	\label{fig:domain_comparison}
\end{figure}

We compare the conservatism of our proposed method \texttt{SLS-MPC} using the full filter parameterization and the baseline robust MPC methods for varying values of $(\epsilon_A, \epsilon_B, \sigma_w)$ and the horizon. For each fixed $(\epsilon_A, \epsilon_B, \sigma_w)$, we first apply an iterative algorithm~\citep{grieder2003robust} to find the maximal robust control invariant set and use it as the terminal set $\mathcal{X}_T$. By construction, $\mathcal{X}_T$ gives the largest feasible domain for any robust MPC algorithm and can be achieved by the exact yet computationally expensive dynamic programming approach~\citep[Chapter 15]{borrelli2017predictive}. We carry out a $10 \times 10$ uniform grid search of initial conditions $x_0$ over $\mathcal{X}_T$ and solve all robust MPC formulations with the sampled $x_0$ using horizon $T = 3$ and $T=10$. The coverage of the feasible domain of each MPC method is given by the ratio of sampled feasible initial conditions and a coverage close to $1$ indicates minimal conservatism of the method even compared with dynamic programming. {In Fig.~\ref{fig:coverage_comparison}, we plot the coverages of the aforementioned robust MPC methods under varying $(\epsilon_A, \epsilon_B, \sigma_w)$ values for horizon $T = 3$ and $T =10$, respectively. In Fig.~\ref{fig:domain_comparison}, we plot the feasible domain of each robust MPC method under the parameter $\epsilon_A = 0.4, \epsilon_B = 0.1, \sigma_w = 0.1$ and horizon $T = 3$ on a $20 \times 20$ grid of initial conditions. The feasible domain is estimated as the convex hull of the feasible initial conditions of each method. }

Fig.~\ref{fig:coverage_comparison} shows that \texttt{SLS-MPC} consistently outperforms all other methods for a wide range of uncertainty parameters and Fig.~\ref{fig:domain_comparison} provides a visual illustration. In all cases \texttt{SLS-MPC} achieves a coverage greater than $90\%$.  {Importantly, from Fig.~\ref{fig:coverage_comparison} we observe that the conservatism of \texttt{SLS-MPC} remains almost unaffected as the horizon increases from $T=3$ to $T=10$ while there is clear conservatism deterioration for all other methods.  For example, the coverage of \texttt{Offline-Tightening-MPC} is similar to those of the tube-based methods with horizon $T=3$ but drops considerably with horizon $T=10$ (see Appendix~\ref{sec:offline_tightening_MPC} for a detailed explanation). }

\begin{remark}
{Although we do not provide a recursive feasibility guarantee with a fixed horizon for \texttt{SLS-MPC}, Fig.~\ref{fig:domain_comparison} and Fig.~\ref{fig:coverage_comparison} suggest that in practice the feasible domain of \texttt{SLS-MPC} with a fixed horizon can be large enough to overcome the effect of the shrinking horizon strategy discussed in Section~\ref{sec:cl_properties}.}
\end{remark}

\subsubsection{Randomly generated examples}
\label{sec:random_example}
\begin{figure}[htb]
	\centering
	\includegraphics[width= 0.6 \columnwidth]{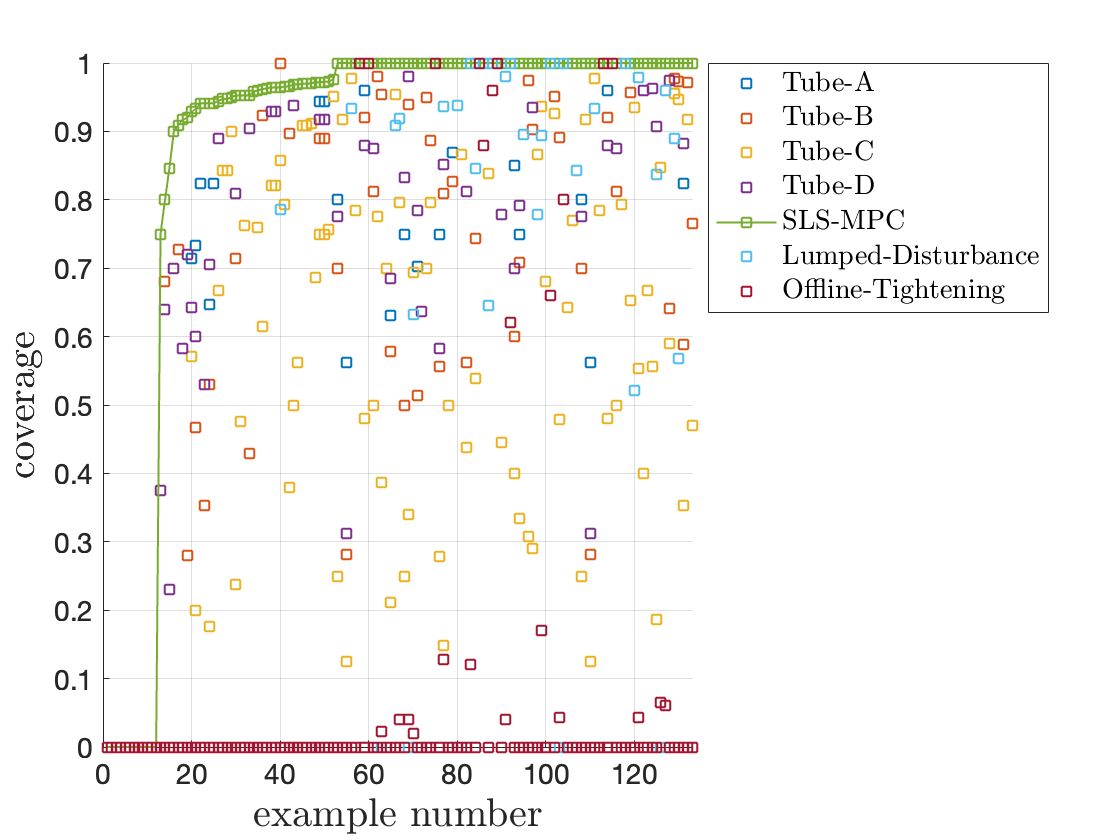}
	\caption{{Coverage comparison of robust MPC methods with horizon $T=10$ on randomly generated examples.}}
	\label{fig:random_comparison}
\end{figure}

{Following the problem setup in Section~\ref{sec:test_example} with horizon $T=10$, we fix the uncertainty parameters $\epsilon_A = 0.2, \epsilon_B = 0.1, \sigma_w = 0.2$. Then, we randomly generate nominal dynamics $\hat{A}, \hat{B}$ by sampling their entries uniformly from a bounded interval and scaling the spectral radius of $\hat{A}$ to a random number in $[0.5, 2.5]$. In Fig.~\ref{fig:random_comparison} we plot the coverages of different robust MPC methods over $130$ randomly generated examples, arranged in an ascending order of the coverages of \texttt{SLS-MPC}. We observe that \texttt{SLS-MPC} achieves the best coverage in almost all examples while no method can consistently achieve the second-best coverage. For a large portion of the randomly generated examples, the coverage gap between \texttt{SLS-MPC} and the second-best method is significant.}


\subsubsection{Effects of tube cross-sections}
\label{sec:tube_cross_section}
In our experiments, we note that the design of the tube cross-section can significantly affect the performance of tube-based MPC. Therefore, for each problem setup, we solved tube-based MPC with three common choices of tube cross-sections in the literature: the minimal robust forward invariant set~\citep{rakovic2005minimal}, the maximal robust forward invariant set~\citep{pluymers2005efficient}, the $\lambda$-contractive set~\citep[Chapter 5.6]{rakovic2016model}. These methods require first computing a robustly stabilizing controller $u_t = Kx_t$ for the uncertain system $x_{t+1} = (\hat{A}+ \Delta_A) x_t + (\hat{B} + \Delta_B) u_t$ with $(\Delta_A, \Delta_B) \in \mathcal{P}$, which can be achieved by solving an SDP~\citep{boyd1994linear}. We choose such $K$ as the pre-stabilizing feedback gain for tube-based MPC and use it to find the tube cross-section. 

\footnotetext{The feasible domain of \texttt{Lumped-Disturbance-MPC} is empty in this example and therefore is not plotted.}

\begin{figure*}
	\centering
	\begin{subfigure}[b]{0.24\textwidth}
		\centering
		\includegraphics[width=\textwidth]{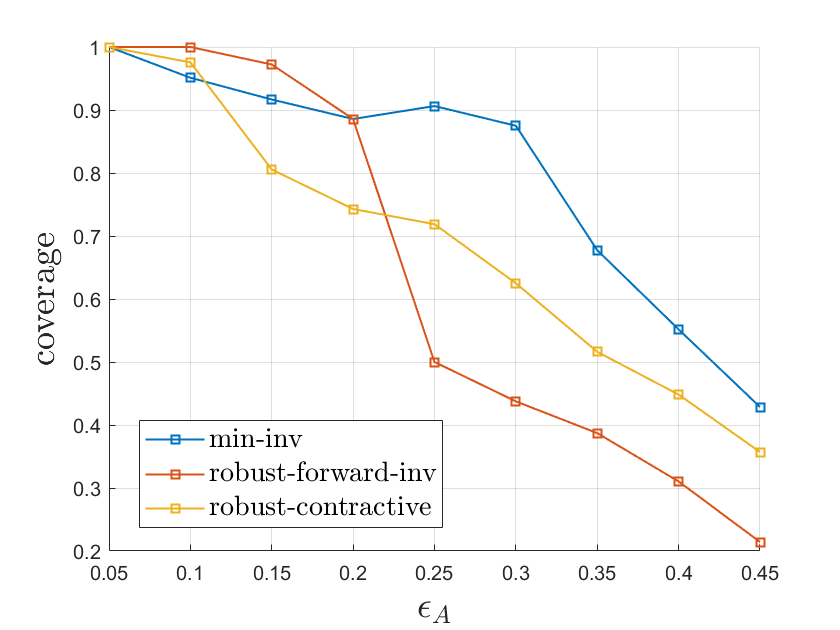}
		\caption{\texttt{Tube-A}}    
		\label{fig:coverage_tube_A}
	\end{subfigure}
	\hfill
	\begin{subfigure}[b]{0.24\textwidth}  
		\centering 
		\includegraphics[width=\textwidth]{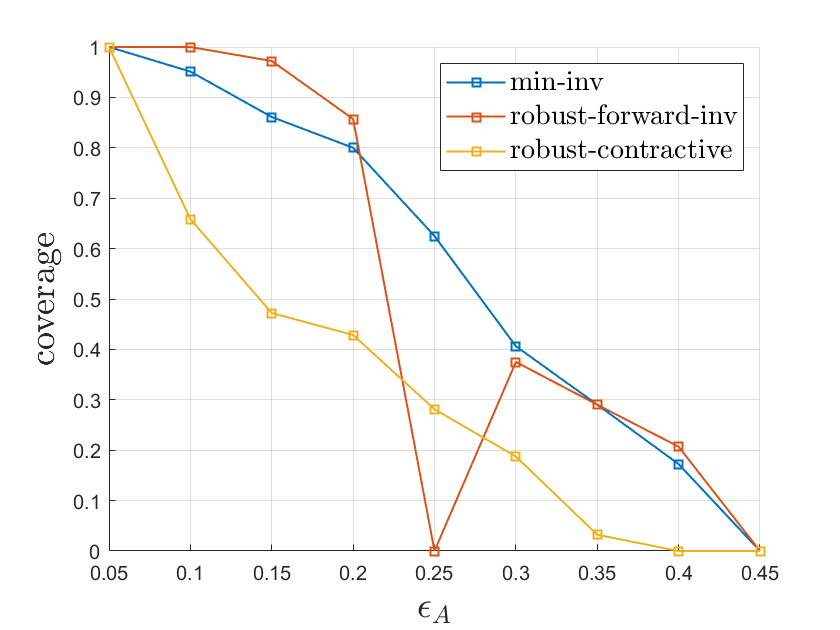}
		\caption{\texttt{Tube-B}}    
		\label{fig:coverage_tube_B}
	\end{subfigure}
\hfill
	\begin{subfigure}[b]{0.24\textwidth}   
		\centering 
		\includegraphics[width=\textwidth]{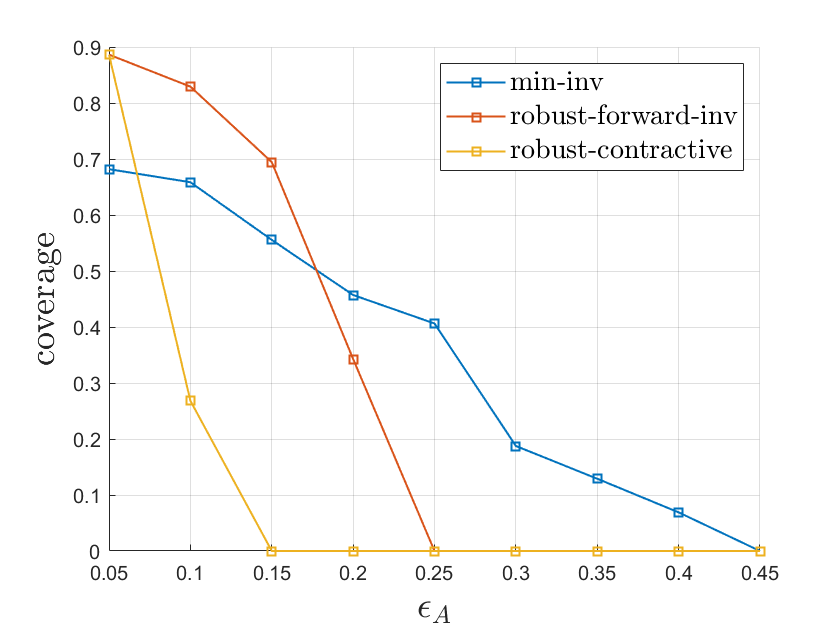}
		\caption{\texttt{Tube-C}}    
		\label{fig:coverage_tube_C}
	\end{subfigure}
	\hfill
	\begin{subfigure}[b]{0.24\textwidth}   
		\centering 
		\includegraphics[width=\textwidth]{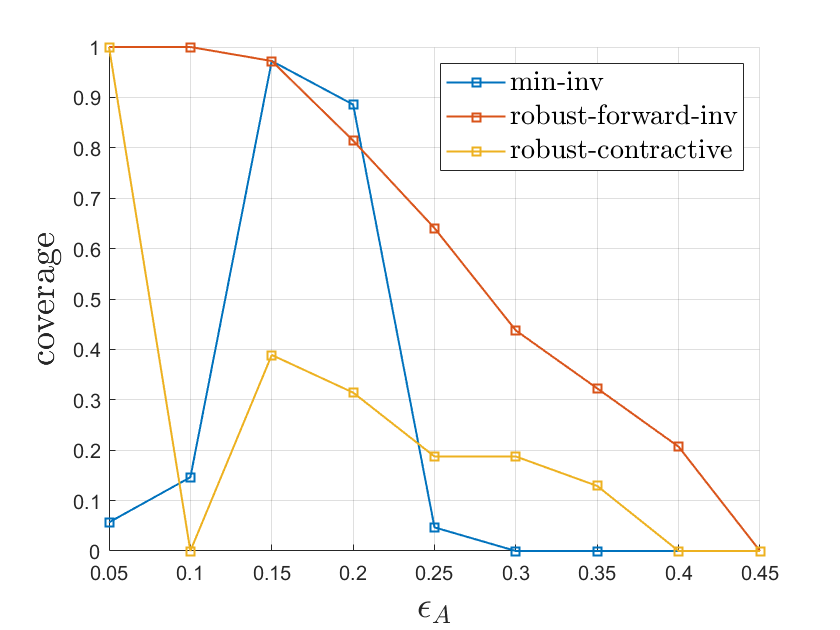}
		\caption{\texttt{Tube-D}}    
		\label{fig:coverage_tube_D}
	\end{subfigure}
	\caption{The coverage of tube-based robust MPC methods with different choices of the tube cross-section. The tube cross-section can be chosen as the minimal nominal forward invariant set (min-inv), maximal robust forward invariant set (robust-foward-inv), or the maximal robust contractive set (robust-contractive). See Section~\ref{sec:tube_cross_section} for details. The coverage is reported for $\epsilon_B = 0.1, \sigma_w = 0.1$ and varying sizes of $\epsilon_A$. } 
	\label{fig:coverage_tube}
\end{figure*}

Fig.~\ref{fig:coverage_tube} demonstrates the dependence of the conservatism of tube-based robust MPC methods on the choice of the tube cross-section. We observe that (i) the choice of the cross-section can greatly affect the conservatism, and (ii) there is no particular choice that consistently performs better than the others for different levels of uncertainties. Importantly, improperly chosen tube cross-sections may cause tube-based MPC to be overly conservative even when the uncertainty level is small, as demonstrated by the cases of $\epsilon_A = 0.1$ in Fig.~\ref{fig:coverage_tube_D} and $\epsilon_A = 0.25$ in Fig.~\ref{fig:coverage_tube_B}. Therefore, in practice it requires manual tuning to figure out which cross-section to use, and the computation of the aforementioned tube cross-sections can be challenging when the system dimension increases. In the feasible domain comparison shown in Fig.~\ref{fig:coverage_comparison} and Fig.~\ref{fig:random_comparison}, for each tube-based method, we report the \emph{best coverage} out of all three tube cross-sections. 

\subsubsection{Evaluation of filter parameterization in \texttt{SLS-MPC}}
\label{sec:filter_comparison}
\begin{figure}
	\centering
	\begin{subfigure}{0.49 \columnwidth}
		\includegraphics[width = 0.9 \textwidth]{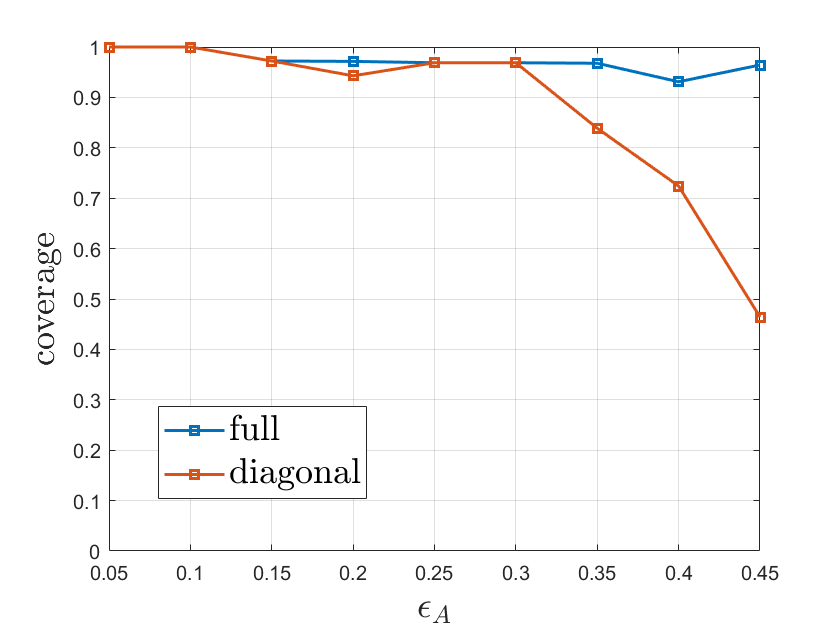}
		\caption{$\epsilon_B = 0.1, \sigma_w = 0.1$}
		\label{fig:filter_eps_A_comparison}
	\end{subfigure}
	\hfill
	\begin{subfigure}{0.49 \columnwidth}
		\includegraphics[width = 0.9 \textwidth]{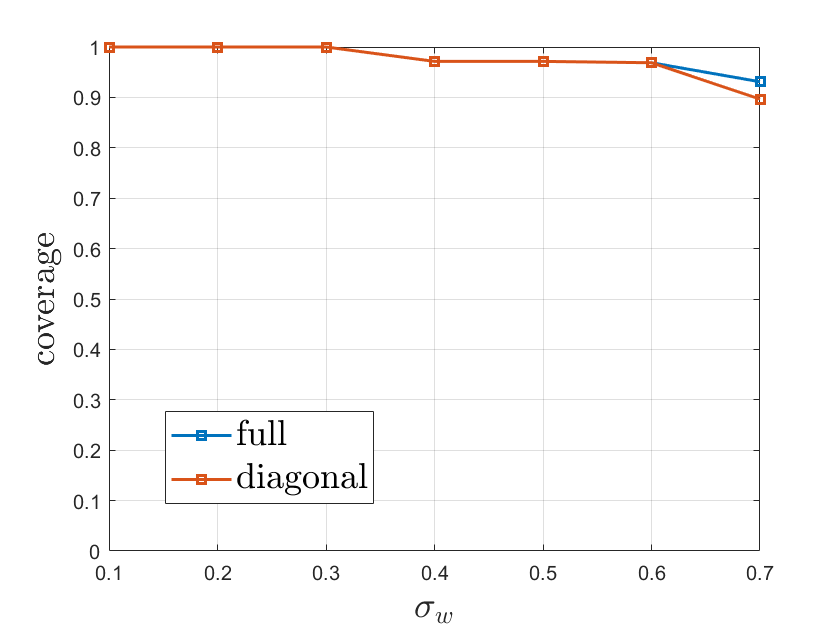}
		\caption{ $\epsilon_A = 0.1, \epsilon_B = 0.1$.}
		\label{fig:filter_w_comparison}
	\end{subfigure}
	\caption{Feasible domain coverage of \texttt{SLS-MPC} using the full (blue) and diagonal (orange) parameterization of the filter for different uncertainty parameters.}
	\label{fig:filter_comparison}
\end{figure}

We now investigate how the parameterization of the filter $\mathbf{\Sigma}$ affects the conservatism of \texttt{SLS-MPC}. With the full parameterization, the filter $\mathbf{\Sigma}$ is a block-lower triangular matrix while with the diagonal parameterization, all off-diagonal blocks are set zero. Fig.~\ref{fig:filter_comparison} shows that the full parameterization of the filter reduces the conservatism of \texttt{SLS-MPC}. When either the model uncertainty or the additive disturbance becomes large, the diagonal parameterization of the filter is not sufficient to mitigate the effects of uncertainties. The average solver time of \texttt{SLS-MPC} with the full (diagonal) parameterization is $0.0703 (0.0601)$ seconds in Fig.~\ref{fig:filter_eps_A_comparison} and $0.0764 (0.0685)$ seconds in Fig.~\ref{fig:filter_w_comparison}.

\begin{figure}
	\centering
	\begin{subfigure}{0.99 \columnwidth}
		\includegraphics[width = \textwidth]{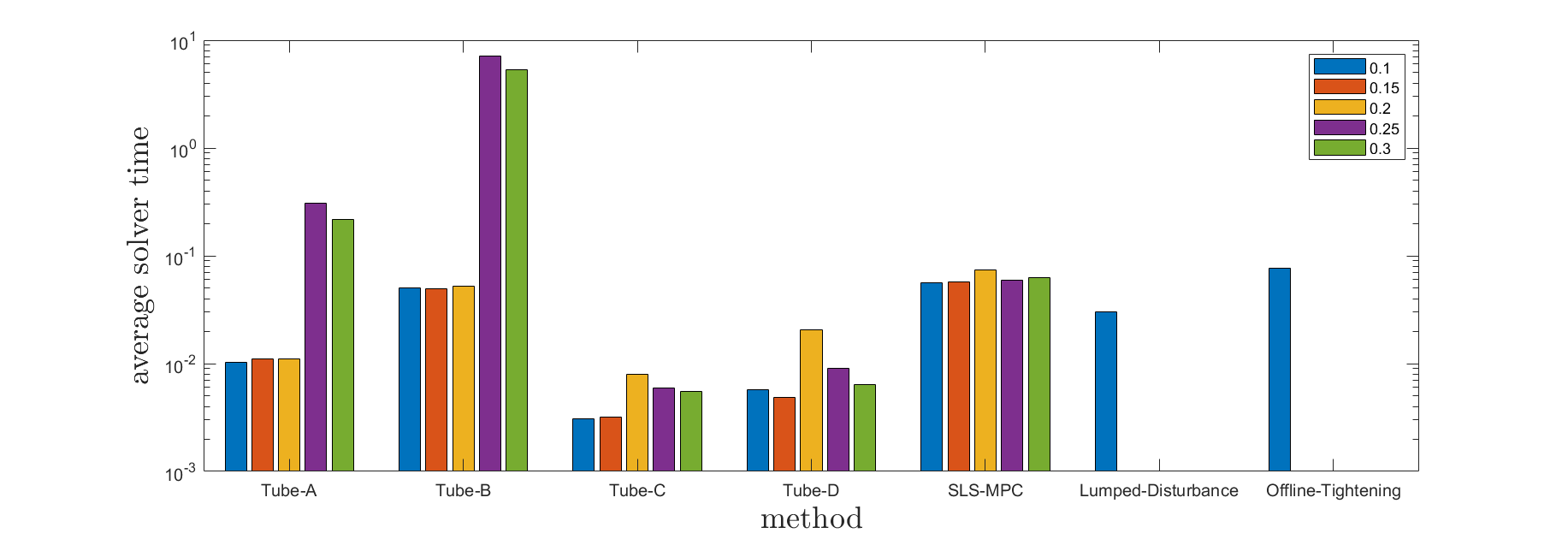}
		\caption{Average solver time of each robust MPC method with horizon $T=10$, $\epsilon_B = 0.1, \sigma_w = 0.1$ and varying values of $\epsilon_A$ as labeled in the legend.}
	\end{subfigure}
	\hfill
	\begin{subfigure}{0.99 \columnwidth}
		\includegraphics[width = \textwidth]{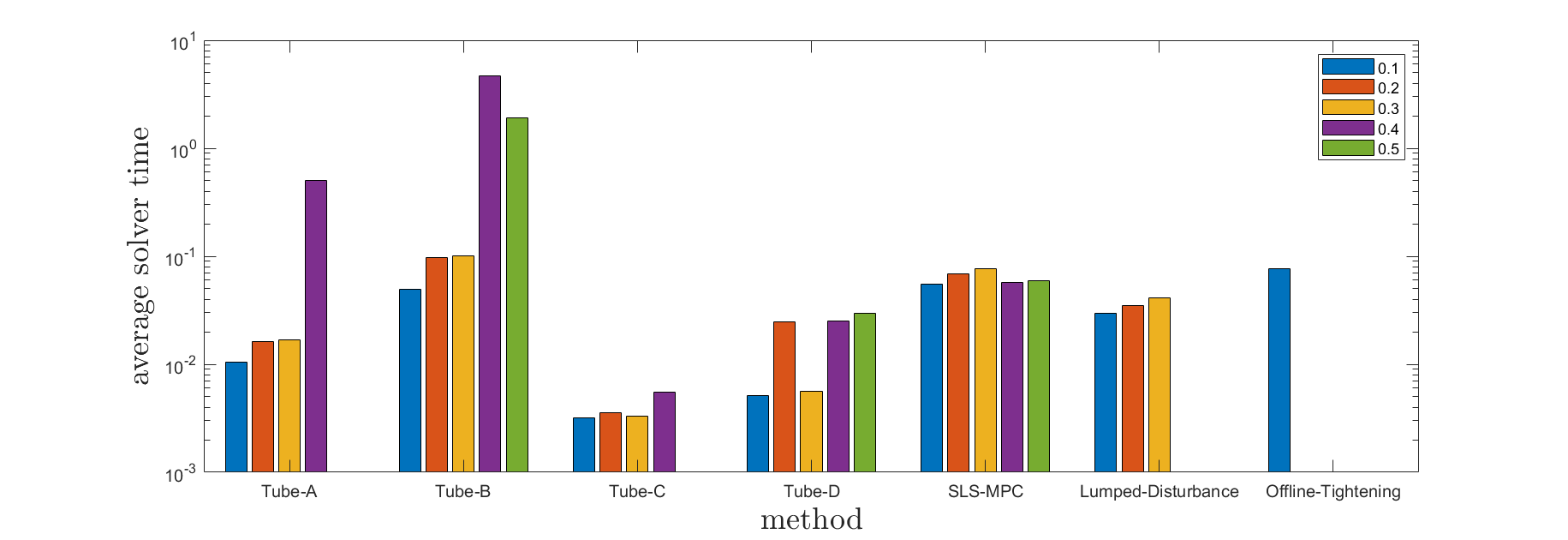}
		\caption{Average solver time of each robust MPC method with horizon $T=10$, $\epsilon_A = 0.1, \epsilon_B = 0.1$ and varying values of $\sigma_w$ as labeled in the legend.}
	\end{subfigure}
	\caption{Average solver time of different robust MPC methods with horizon $T=10$ in the coverage evaluation (see Fig.~\ref{fig:coverage_comparison}) under different uncertainty parameters. }
	\label{fig:solver_time_comparison}
\end{figure}

\subsection{Solver time comparison}
In Fig.~\ref{fig:solver_time_comparison}, we plot the average solver time of each robust MPC method in the conservatism comparison experiment shown in Section~\ref{sec:feasible_domain_comparison} with varying uncertainty parameters~\footnote{For tube-based MPC, the cross-section that gives the best coverage is used to generate this figure.} and horizon $T=10$. Data is left blank if a robust MPC method was not feasible on any sampled initial conditions. We observe that the solver time of \texttt{Tube-C} and \texttt{Tube-D} is normally an order of magnitude smaller than that of the other methods. The average solver times of the uncertainty over-approximation-based methods \texttt{SLS-MPC}, \texttt{Lumped-Disturbance-MPC}, \texttt{Offline-Tightening-MPC} are close since these methods all search for an LTV state feedback controller. We observe that the solver time of robust MPC methods can significantly increase when a challenging problem instance (e.g., those with large uncertainty parameters) is encountered. 

\section{Conclusion}
\label{sec:conclusion}
We proposed a novel robust MPC method for uncertain linear systems subject to both polytopic model uncertainty and additive disturbances. Using System Level Synthesis, our method searches for a robust LTV state feedback controller in the space of closed-loop system responses which allows efficient over-approximation of the effects of uncertainty and optimization over upper bounds on the worst-case costs with respect to the model uncertainty.  Numerical examples demonstrate that our method can significantly reduce the conservatism compared with a wide range of tube-based and uncertainty over-approximation-based robust MPC methods. 

\appendix
\section{Appendix}
\label{sec:app}

\subsection{Proof of Lemma~\ref{lem:equivalence}}
\label{app:proof_lemma}
As commented in Section~\ref{sec:eta_dynamics}, for any realization of $\{\DDelta_A, \DDelta_B, \ww\}$, the values of the lumped uncertainty $\bfeta$ are given by~\eqref{eq:eta_dynamics} and uniquely defined. We first assume Eq.~\eqref{eq:set_over_approx}, i.e., $\mathcal{R}(\bfeta;\{\Phix, \Phiu\}) \subseteq \mathcal{R}(\bfSigma \tildeww)$ holds. This means for any realization of uncertainty and the induced lumped uncertainty values $\bfeta^\star$, we can always find $\tildeww^\star \in \mathcal{W}_{\tildeww}$ such that $\bfeta^\star = \bfSigma \tildeww^\star$. Plugging $\tildeww^\star$ into~\eqref{eq:eta_dynamics} readily gives that $\tildeww^\star$ is a solution to~\eqref{eq:robust_equalities}. Therefore, condition~\eqref{eq:set_over_approx} indicates that Eq.~\eqref{eq:robust_equalities} holds robustly. 

Next, we assume Eq.~\eqref{eq:robust_equalities} holds robustly. Then, for any realization of $\{\DDelta_A, \DDelta_B, \ww\}$, let $\tildeww^\star$ denote the solution to~\eqref{eq:robust_equalities} and define $\boldsymbol{\xi}^\star = \bfSigma \tildeww^\star$. Let $\bfeta^\star$ denote the solution to~\eqref{eq:eta_dynamics} with the same uncertainty parameters. Due to the uniqueness of the solution, we have $\bfeta^\star = \boldsymbol{\xi}^\star = \bfSigma \tildeww^\star$. Therefore, for any $\bfeta^\star \in \mathcal{R}(\bfeta; \{\Phix, \Phiu\})$, we can always find $\tildeww^\star \in \mathcal{W}_{\tildeww}$ such that $\bfeta^\star = \bfSigma \tildeww^\star$, and hence \eqref{eq:set_over_approx} holds. 

\subsection{Proof of Corollary~\ref{coro:scaled_equivalence}}
\label{app:proof_corollary}
1. For a given controller $\KK$, similar to~\eqref{eq:explicit_map}, the system responses from $\tildeww$ to $(\xx, \uu)$ are given by $\tildePhix = (I - Z(\hat{\sA} + \hat{\sB} \KK))^{-1} \bfSigma$, $\tildePhiu = \KK (I - Z(\hat{\sA} + \hat{\sB} \KK))^{-1} \bfSigma$ and satisfy the affine constraint~\eqref{eq:affine_scaled}.

2. First note that the block diagonal of $\tildePhix$ and $\bfSigma$ are equal according to~\eqref{eq:affine_scaled}. Then, $\bfSigma$ being invertible indicates that $\tildePhix^{-1}$ exists. For any $\{\tildePhix, \tildePhiu\}$ satisfying~\eqref{eq:affine_scaled}, we synthesize the state feedback controller as $\KK = \tildePhiu \tildePhix^{-1}$ which satisfies $I - Z\hat{\sA} -Z \hat{\sB} \KK = \bfSigma \tildePhix^{-1}$ by multiplying equation~\eqref{eq:affine_scaled} with $\tildePhix^{-1}$ from right. Then we have $(I - Z(\hat{\sA} + \hat{\sB} \KK))^{-1} \bfSigma = \tildePhix \bfSigma^{-1} \bfSigma = \tildePhix$ and $\KK (I - Z(\hat{\sA} + \hat{\sB} \KK))^{-1} \bfSigma = \tildePhiu \tildePhix^{-1} \tildePhix = \tildePhiu$.

\subsection{Robust MPC baselines}
\label{sec:mpc_baselines}
Here we explain the features and implementation of the robust MPC baselines used in Section~\ref{sec:simulation}. These methods mainly vary in (i) how to parameterize the control policy $\pi_t$ and (ii) how to handle the polytopic model uncertainty such that the state and input constraints are robustly satisfied under the synthesized policy $\pi_t$. 

\subsubsection{Tube-based robust MPC}
Tube-based methods, which aim to synthesize a robust controller and an associated tube to contain all possible realization of system trajectories, are popular in robust MPC. Tube-based MPC typically follows the following scheme:
\begin{enumerate}
	\item Parameterize the control policy $\pi_t$.
	\item Parameterize the tube cross-sections $\mathbb{X}_t$.
	\item Enforce the tube containment constraints inductively:
	\begin{equation} \label{eq:tube_containment}
		\begin{aligned}
		x_t \in \mathbb{X}_t \Rightarrow x_{t+1} \in \mathbb{X}_{t+1}, \ \forall (\Delta_A, \Delta_B) \in \mathcal{P}, \forall w_t \in \mathcal{W}, t = 0, \cdots, T-1.
		\end{aligned}
	\end{equation}
	and guarantee robust satisfaction of all state and input constraints:
	\begin{equation} \label{eq:tube_robustness}
		\begin{aligned}
			\mathbb{X}_t \subseteq \mathcal{X}, u_t = \pi_t(x_t) \in \mathcal{U}, \ \forall x_t \in \mathbb{X}_t, t = 0, \cdots, T-1, \ x_0 \in \mathbb{X}_0, \ \mathbb{X}_T \in \mathcal{X}_T.
		\end{aligned}
	\end{equation}
	\item Specify the cost function and solve a robust OCP with constraints~\eqref{eq:tube_containment} and~\eqref{eq:tube_robustness}.
\end{enumerate}
Steps $1$ to $3$ in the above scheme jointly determine the conservatism and computational complexity of a tube-based robust MPC method. Next, we review all four tube-based robust MPC baselines shown in Table~\ref{table:tube}. 

\paragraph{Tube-A}
The first method which we denote \texttt{Tube-A} is from~\citet{langson2004robust}. In \texttt{Tube-A}, a {homothetic} tube~\citep{rakovic2012homothetic} is applied:
\begin{equation} \label{eq:homo_tube}
	\mathbb{X}_t = z_t + \alpha_t \mathbb{X}, t = 1, \cdots, T
\end{equation}
where $z_t \in \mathbb{R}^{n_x}$ and $\alpha_t \geq 0$ is a scalar. The tube cross-sections $\mathbb{X}_t$ at different time instants are restricted to be transitions and dilations of a given polytope $\mathbb{X}$. 

Let $n_J$ denote the number of vertices of $\mathbb{X}$ and $\{x^j\}_{j=1}^{n_J}$ denote the vertices of $\mathbb{X}$. Since $\mathbb{X}_t$ has the same number of vertices as $\mathbb{X}$, we can define $\{x_t^j\}_{j=1}^{n_J}$ as the vertices of $\mathbb{X}_t$ similarly. For controller design, \texttt{Tube-A} associates a control input $u_t^j$ with each vertex $x_t^j$ of $\mathbb{X}_t$, and parameterize the feedback policy $\pi_t$ in a barycentric manner:
\begin{equation}\label{eq:tube_mpc_controller}
	\begin{aligned}
		u_t = \pi_t(x_t) = \sum_{j = 1}^{n_J} \lambda_j u_t^j, \text{where $\lambda_j$ satisfies } x_t = \sum_{j=1}^{n_J} \lambda_j x_t^j, \lambda_j \geq 0, \sum_{j=1}^{n_J} \lambda_j = 1.
	\end{aligned}
\end{equation}
With the homothetic tube and barycentric controller parameterization, constraints~\eqref{eq:tube_containment} and~\eqref{eq:tube_robustness} can be enforced by only considering the vertex states $x_t^j$ and control inputs $u_t^j$. In this process, both the vertices of the tube cross-section $\mathbb{X}_t$ and the polytopic model uncertainty set $\mathcal{P}$ are enumerated.


\paragraph{Tube-B}
In~\citet{lorenzen2019robust}, the homothetic tube parameterization~\eqref{eq:homo_tube} is applied with a pre-stabilizing controller $\pi_t(x_t) = Kx_t + v_t$. We denote this method as \texttt{Tube-B} where the time-invariant feedback gain $K$ is found offline by solving an SDP such that $u_t = Kx_t$ is robustly stabilizing for the uncertain system~\eqref{eq:dyn} with polytopic model uncertainty. The bias terms $v_t$ of the control inputs are optimized online together with the center $z_t$ and scaling parameter $\alpha_k$ of the cross-sections. Constraints~\eqref{eq:tube_containment} and~\eqref{eq:tube_robustness} are enforced using a dual formulation and enumeration of the vertices of each cross-section $\mathbb{X}_t$.

\paragraph{Tube-C}
\citet{kohler2019linear} propose a tube-based robust MPC method that applies a homothetic tube~\eqref{eq:homo_tube} and a pre-stabilizing feedback controller $u_t = Kx_t + v_t$. Compared with \texttt{Tube-B}, this method bounds the effects of uncertainty on the dilation of the tube cross-sections offline instead of using vertex enumeration of the tube cross-section.  Furthermore, in~\citet{kohler2019linear} the centers of the cross-sections $\mathbb{X}_t$ are chosen as the nominal trajectory $\hat{x}_t$ evolved according to $\hat{x}_{t+1} = \hat{A}\hat{x}_t + \hat{B}(K\hat{x}_t + v_t)$; in contrast, in \texttt{Tube-A} and \texttt{Tube-B}, the centers $z_t$ of $\mathbb{X}_t$ are treated as free variables. We label the robust MPC method proposed in~\citet{kohler2019linear} as \texttt{Tube-C}.

\paragraph{Tube-D}
We denote the robust MPC method proposed in~\citet{lu2019robust} as \texttt{Tube-D}. It uses the pre-stabilizing feedback controller $u_t = Kx_t + v_t$ but parameterizes the cross-sections as $\mathbb{X}_t = \{x \in \mathbb{R}^{n_x} \mid V x \leq \beta_t \}$, where $V \in \mathbb{R}^{r \times n_x}$ denotes a fixed set of hyperplanes that define a polytope and the hyperplane offsets $\beta_t \in \mathbb{R}^r$ are optimized online. This formulation of $\mathbb{X}_t$ is more flexible than the homothetic tube~\eqref{eq:homo_tube}, but treating $\beta_t$ as optimization variables give rise to bilinear terms of the form $\Lambda \beta_t$ in enforcing constraints~\eqref{eq:tube_containment} and~\eqref{eq:tube_robustness}, where $\Lambda$ denotes the dual variables in a Lagrangian function. To avoid numerical intractability, the dual variables $\Lambda$ are fixed offline by solving a set of linear programs. 

\subsubsection{Uncertainty over-approximation-based MPC}
\label{app:state_feedback_MPC}
For the uncertain system $x_{t+1} = \hat{A}x_t + \hat{B}u_t + w_t$ with only additive disturbances, \citet{goulart2006optimization} prove that the LTV state feedback controller 
$\uu = \mathbf{L} \xx + \mathbf{g}$ is equivalent to the disturbance feedback controller $\uu = \mathbf{M}\ww + \mathbf{v}$~\footnote{We refer the readers to \citet{goulart2006optimization} for the exact parameterization of $\mathbf{L}, \mathbf{g}, \mathbf{M}, \mathbf{v}$.}. 
To extend the state/disturbance feedback approach to handle both the polytopic model uncertainty and the additive disturbances, the effects of uncertainty have to be properly over-approximated for the robust OCP to be numerically tractable. 

\paragraph{Lumped-Disturbance-MPC}
A simple extension of state/disturbance feedback controllers to handle model uncertainties is to uniformly over-approximate the perturbation $\Delta_A x_t + \Delta_B u_t + w_t$ by a norm-bounded disturbance $\bar{w}_t$, and apply standard state/disturbance feedback controller design on the system $x_{t+1} = \hat{A}x_t + \hat{B}u_t + \bar{w}_t$. Since the state constraint $\mathcal{X}$ and input constraint $\mathcal{U}$ are compact sets, an upper bound on $\lVert \Delta_A x_t + \Delta_B u_t + w_t \rVert_\infty$ can be easily obtained using the triangle inequality and submultiplicativity of norms as shown in~\citet{bujarbaruah2021simple}. We denote this method proposed in~\citet{bujarbaruah2021simple} as \texttt{Lumped-Disturbance-MPC}. Despite being simple, the uniform norm-bounded uncertainty over-approximation can be conservative since it is agnostic to the current state of the system. Our proposed method, SLS MPC, tries to resolve this issue by maintaining the dependence of the uncertainty over-approximation on the system states.

\paragraph{Offline-Tightening-MPC}
\label{sec:offline_tightening_MPC}
\citet{bujarbaruah2022robust} propose a constraint tightening method for the robust OCP~\eqref{eq:robustOCP} such that the disturbance feedback controller parameters $\mathbf{M}$ and $\mathbf{v}$ explicitly control the tightening margin. However, the constraint tightening parameters also heavily rely on offline computed relaxation bounds that hold for all possible open-loop control inputs, which could already become conservative before the disturbance feedback controller is plugged in. We denote this method from~\citet{bujarbaruah2022robust} as \texttt{Offline-Tightening-MPC}. In the implementation of this method, the constraint tightening margins are computed offline using a hybrid approach of uncertainty set vertex enumeration and norm bounds over-approximation (see~\citet[Appendix A.4]{bujarbaruah2020robust}). Due to the combinatorial complexity, we only enumerate the vertices of the uncertainty set~\eqref{eq:uncertainty_form} for a truncation horizon of $N = 3$ and use norm bounds over-approximation for the rest $T-N$ predictive steps. Therefore, the conservatism of such offline computed margins increases as the MPC horizon $T$ becomes large. In our experiments, we use $T = 10$ (see Section~\ref{sec:simulation_conservatism}) which is much larger than the tested horizon $T = 1,2,3$ used in the simulation examples in~\citet{bujarbaruah2022robust}. 



\subsection{Additional conservatism comparison results}
\label{sec:original_example}
{In this section, we compare the coverages of all robust MPC methods for the example from~\citet{bujarbaruah2022robust} where the model uncertainty is given by}
\begin{equation} \label{eq:original_model_uncertainty}
	\begin{aligned}
		&\Delta_A \in \text{Conv} \Big \{ \begin{bmatrix}
			0 & \pm \epsilon_A \\ \pm \epsilon_A & 0
		\end{bmatrix}  \Big \}, \\
	  &\Delta_B \in \text{Conv}\Big \{ \begin{bmatrix}
			0 \\ \pm \epsilon_B
		\end{bmatrix},  \begin{bmatrix}
			\pm \epsilon_B \\ 0
		\end{bmatrix} \Big \},
	\end{aligned}
\end{equation}
{with $\epsilon_A = 0.1, \epsilon_B = 0.1,$ and the additive disturbances are bounded by $\lVert w_t \rVert_\infty \leq 0.1$.  This model uncertainty set has $16$ vertices in the joint space of $(\Delta_A, \Delta_B)$, for which the maximal robust control invariant set becomes empty at $\epsilon_A = 0.14$. This prevents us from comparing different methods over a range of uncertainty parameters.}

{In Table~\ref{tab:original}, we report the coverages of the robust MPC methods with horizon $T = 3$ and $T = 10$. For each tube-based MPC method, we report the coverage by $(c_1, c_2, c_3)$ where $c_1, c_2, c_3$ denote the coverage obtained with the minimal robust forward invariant set, the maximal robust forward invariant set, and the $\lambda$-contractive set as the tube cross-sections, respectively (see Section~\ref{sec:tube_cross_section} for details). The best coverage is highlighted in boldface. We observe that the coverage of \texttt{SLS-MPC} is among the best and the conservatism of the tube-based methods depends on the cross-section used.  These observations are consistent with what was shown in Section~\ref{sec:simulation_conservatism}.}

\begin{table*}[tb]
	\caption{{Coverage comparison of robust MPC methods with horizon $T = 3$ and $T=10$ for the example from~\citet{bujarbaruah2022robust}.  The $3$-tuple denotes the coverages of tube-based methods using different cross-sections.}}
	\label{tab:original}
	\resizebox{\textwidth}{!}{
		\begin{tabular}{|c|c|c|c|c|c|c|c|} 
			\toprule
			Method  & \texttt{Tube-A} & \texttt{Tube-B} & \texttt{Tube-C} & \texttt{Tube-D}& \texttt{SLS-MPC} & \texttt{Lumped-Disturbance} & \texttt{Offline-Tightening}  \\ \midrule 
			Coverage (hor. $=3$) & $(0.66, \mathbf{1}, 1)$ & $(0, \mathbf{1}, 1)$ & $(0.12, 0.78, \mathbf{0.88})$  &  $(0.95, \mathbf{1}, 1)$ & $0.98$       & $ 0.80$  &  $ 0.85$\\  \midrule
			Coverage (hor.$=10$) & $(0.34, \mathbf{1}, 1)$ & $(0, \mathbf{1}, 0.88)$ & $(0, \mathbf{0.39}, 0.05)$  &  $(0.93, \mathbf{1}, 0.89)$ &  $0.98 $      & $0.49 $  &  $0.02$\\ 
			\bottomrule
		\end{tabular}
	}
\end{table*}

\bibliographystyle{apalike}
\bibliography{Refs}

\end{document}